\documentclass[prl, aps,reprint,groupedaddress,twocolumn,amsmath,amssymb,floatfix]{revtex4-2}

\usepackage[colorlinks,linkcolor=blue,anchorcolor=blue,citecolor=blue]{hyperref}
\usepackage{graphicx}
\usepackage{braket} 
\usepackage{subfigure}
\usepackage{floatrow}
\usepackage{dcolumn}
\usepackage{bm}
\usepackage{autobreak}
\usepackage{url}
\usepackage{booktabs} 

\usepackage{makecell}

\begin{document}


\title{
A Control Line Multiplexing Framework for Scalable Quantum Computing
}


\author{Dayu~Li$^{1,2,3}$}
\author{Naibin~Zhou$^{1,2,3}$}
\author{Zhiyuan~Chen$^{1,2,3}$}
\author{Zihua~Chen$^{1,2,3}$}
\author{Yu~Xu$^{3}$}
\author{Yuan~Li$^{1,2,3}$}
\author{Daojin~Fan$^{3}$}
\author{Wenhao~Chu$^{5}$}
\author{Xiawei~Chen$^{3}$}
\author{Hui~Deng$^{1,2,3}$}
\author{Jiacheng~Gui$^{3}$}
\author{Linyin~Hong$^{5}$}
\author{Bo~Fan$^{7}$}
\author{Weiping~Lin$^{1,2,3}$}
\author{Haonan~Zou$^{7}$}
\author{Haoxin~Lou$^{7}$}
\author{Xintao~Zhao$^{7}$}
\author{Sirui~Cao$^{1,3}$}
\author{Ying~Chong$^{3}$}
\author{Qingling~Zhu$^{3}$}
\author{Chen~Wang$^{1,2,3}$}
\author{Yisen~Hu$^{1,2,3}$}
\author{Kaili~Zhang$^{3}$}
\author{Feifan~Su$^{3}$}
\author{Haibin~Zhang$^{3}$}
\author{Jianghan~Yin$^{7}$}
\author{Yajie~Du$^{7}$}
\author{Yinan~Jiang$^{7}$}
\author{Haoping~Luo$^{7}$}
\author{Weifeng~Yang$^{3}$}
\author{Jiafei~Wang$^{5}$}
\author{Jin~Lin$^{2,3}$}
\author{Futian~Liang$^{1,3}$}
\author{Hao~Rong$^{1}$}
\author{Kai~Yan$^{3}$}
\author{Ming~Gong$^{1,2,3}$}
\author{Jinjin~Li$^{3}$}
\author{Maliang~Liu$^{3,6}$}
\author{Cheng-Zhi~Peng$^{1,2,3}$}
\author{Shaowei~Li$^{3}$}
\author{Xiaobo~Zhu$^{1,2,3,4}$}
\author{Jian-Wei~Pan$^{1,2,3}$}

\affiliation{$^1$Hefei National Research Center for Physical Sciences at the Microscale and School of Physical Sciences, University of Science and Technology of China, Hefei 230026, China}
\affiliation{$^2$Shanghai Research Center for Quantum Science and CAS Center for Excellence in Quantum Information and Quantum Physics, University of Science and Technology of China, Shanghai 201315, China}
\affiliation{$^3$Hefei National Laboratory, University of Science and Technology of China, Hefei 230088, China}
\affiliation{$^4$Jinan Institute of Quantum Technology and Hefei National Laboratory Jinan Branch, Jinan 250101, China}
\affiliation{$^5$QuantumCTek Co., Ltd., Hefei 230088, China}
\affiliation{$^6$School of Microelectronics, Xidian University, Xi'an, China}
\affiliation{$^7$University of Science and Technology of China, Shanghai Research Institute, Shanghai 201315, China}

\begin{abstract}
Universal fault-tolerant superconducting quantum computers are expected to require millions of physical qubits; yet extending the prevailing dedicated-control-line paradigm to this scale would exceed the wiring density and cooling-power budget of dilution refrigerators. 
Microwave frequency-division multiplexing (FDM) revolutionized classical communications by enabling many independent channels to share common transmission infrastructure. Translating this principle to superconducting quantum control, however, requires an all-microwave architecture and experimental validation of frequency-multiplexed gate control. Recent work in quantum error correction has demonstrated microwave-based readout, single-qubit gates and leakage suppression~\cite{he2025experimental}, whereas two-qubit gates remain predominantly driven by unipolar pulses and thus constitute the missing element of an all-microwave architecture. Here, we propose a periodically extensible hybrid multiplexing framework in which all gate operations are frequency-division multiplexed using polychromatic microwave pulses and band-pass filters, thereby reducing the required control-line count by one to two orders of magnitude. In a 1:4 demonstration, we achieve average errors of $(6.4\pm0.2)\times10^{-3}$ for parallel FDM CZ gates and $(10.6\pm0.2)\times10^{-4}$ for parallel FDM \(XY(\pi/2)\) gates, without statistically significant multiplexing-induced degradation. Together with on-chip DC sources~\cite{jiang2025chip}, this FDM-based gate-control scheme completes the last missing module in a scalable multiplexed-control architecture toward universal fault-tolerant quantum computing.

\end{abstract}


\maketitle
\section{Introduction}

Realizing universal fault-tolerant quantum computing on superconducting platforms will require processors with very large numbers of physical qubits~\cite{google2025quantum, PhysRevX.12.021012, Chen_2023, gottesman2009introductionquantumerrorcorrection}. 
Factoring a 2048-bit RSA integer, for example, requires hundreds of thousands of physical qubits~\cite{gidney2025factor, mundada2026heterogeneous}. 
Leading experimental systems use approximately three control lines per qubit~\cite{gao2025establishing, he2025experimental, google2025quantum, sivak2025reinforcement}, implying nearly a million external control lines at this scale. 
At this scale, the resulting milliwatt-scale heat load at the mixing-chamber (MXC) plate~\cite{tian2025high} would far exceed the \(5\text{--}30\,\mu\mathrm{W}\) cooling power of typical dilution refrigerators~\cite{tian2025high, krinner2019engineering, raicu2025cryogenic}, even before physical-space constraints and system complexity are considered.
These constraints have motivated efforts to relocate control functionality toward the processor, including cryo-CMOS electronics~\cite{patra2017cryo, pauka2021cryogenic, Chakraborty2022CryoCMOS}, single-flux-quantum logic~\cite{mcdermott2014accurate, li2019hardware, liu2023single}, and optical-fibre links with cryogenic optoelectronic conversion~\cite{Lecocq2021PhotonicLink, Xu2025NullBiasedFiberLink}. Although these approaches can reduce the number of interconnects between room temperature and the cryogenic stages, they rely on active devices or optoelectronic converters at low temperature. The associated power dissipation, together with stringent requirements for noise, bandwidth, integration, reliability and control fidelity, makes it challenging to achieve a net system-level reduction in wiring complexity and MXC heat load. 
By contrast, control-line multiplexing offers a more direct route to alleviating this bottleneck by allowing multiple operations to share each cryogenic interconnect and can leverage comparatively mature technologies, including multitone waveform generation~\cite{ohira2024optimizing}, band-pass filters~\cite{luo2019miniaturized} and multiplexers~\cite{acharya2023multiplexed, Takeuchi2024Microwave}, making it particularly attractive in terms of practical feasibility, engineering cost, and compatibility with existing superconducting platforms.

Universal quantum computing requires single-qubit gates, two-qubit gates, and readout. 
Readout has already benefited from frequency-division multiplexing (FDM)~\cite{yeo2023high, spring2025fast}, whereas qubit and coupler control remain largely unmultiplexed~\cite{gao2025establishing, he2025experimental, google2025quantum, sivak2025reinforcement}. For example, the ZCZ3 platform uses 105 qubit control-lines, 182 coupler-control lines and only 30 readout lines~\cite{gao2025establishing}, showing that gate control constitutes the dominant wiring overhead. 

Several multiplexed control schemes have been proposed to reduce the wiring overhead of single- and two-qubit gates~\cite{shi2023multiplexed, zhao2024multiplexedcontrolarchitecturesuperconducting, tian2025youtiao, mitarai2026orthogonal}, and some have also been experimentally demonstrated~\cite{cai2025multiplexed, matsuda2025selective, asaad2016independent}. Although FDM has been proposed theoretically, experiments have so far demonstrated only TDM for two-qubit gate control, which reduces the number of control lines at the cost of gate parallelism. This restriction arises from the unipolar flux-pulse gates used in these schemes: their short durations enable high fidelity by limiting decoherence exposure, but they lack the carrier-frequency degree of freedom and are therefore not naturally amenable to FDM~\cite{Barends_2014, foxen2020demonstrating, PhysRevApplied.14.024070, PhysRevLett.125.240502, Ye_2021, PhysRevLett.126.220502}. Microwave-based two-qubit gates offer a route around this limitation because each gate can be driven by a single microwave tone~\cite{li2022realization, ma2025analysis, shirai2023all, shirai2025highfidelityallmicrowaveczgate, ding2023high, huang2025fast, kugut2025interaction}, enabling frequency-selective parallel two-qubit control. Combined with parallel FDM single-qubit gates, this capability supplies the entangling primitive needed to extend FDM beyond readout to a universal gate-control architecture. Although the microwave-based two-qubit gates currently require longer durations and are consequently more susceptible to decoherence~\cite{li2022realization, ma2025analysis, shirai2023all, shirai2025highfidelityallmicrowaveczgate}, this trade-off is expected to become less restrictive with advances in qubit design and fabrication~\cite{bland2025millisecond}. These considerations motivate an experimental demonstration of parallel FDM for both single- and two-qubit gates in a superconducting quantum processor. 

Here we establish a scalable multiplexing architecture for gate control and provide a detailed analysis of the operating principles and error mechanisms of the microwave-pulse controlled-Z (MP-CZ) gate used in our experiments. Using simulated crosstalk and error constraints, we evaluate a periodically extensible control-line architecture that combines FDM, TDM and on-chip DC sources (OCDCS)~\cite{jiang2025chip, castellano2006static, deng2015working, bunyk2014architectural, acharya2023multiplexed, hu2026millikelvin}. Using the demonstrated gate parameters, this architecture is projected to reduce the required control-line count to \(1/12\) of the unmultiplexed baseline and to \(1/48\) under the stated idealized assumptions; details of these estimates are provided in Supplementary Section~I. Constrained by the processor size and the available output bandwidth of the control electronics, we demonstrate a 1:4 FDM configuration with average errors of $(10.6\pm0.2)\times10^{-4}$ for parallel single-qubit gates and $(6.4\pm0.2)\times10^{-3}$ for parallel two-qubit gates, the latter constituting the first such demonstration in a superconducting quantum processor. With waveform pre-distortion, neither gate class exhibits a statistically significant error increase relative to its dedicated-channel baseline, indicating that the multiplexing layer introduces no measurable additional error at the present experimental resolution. Together, these results establish the gate-level feasibility of the architecture and support its periodic extrapolation toward scalable fault-tolerant quantum control.


\begin{figure}[htbp]
    \centering 
    \includegraphics[width=\textwidth]{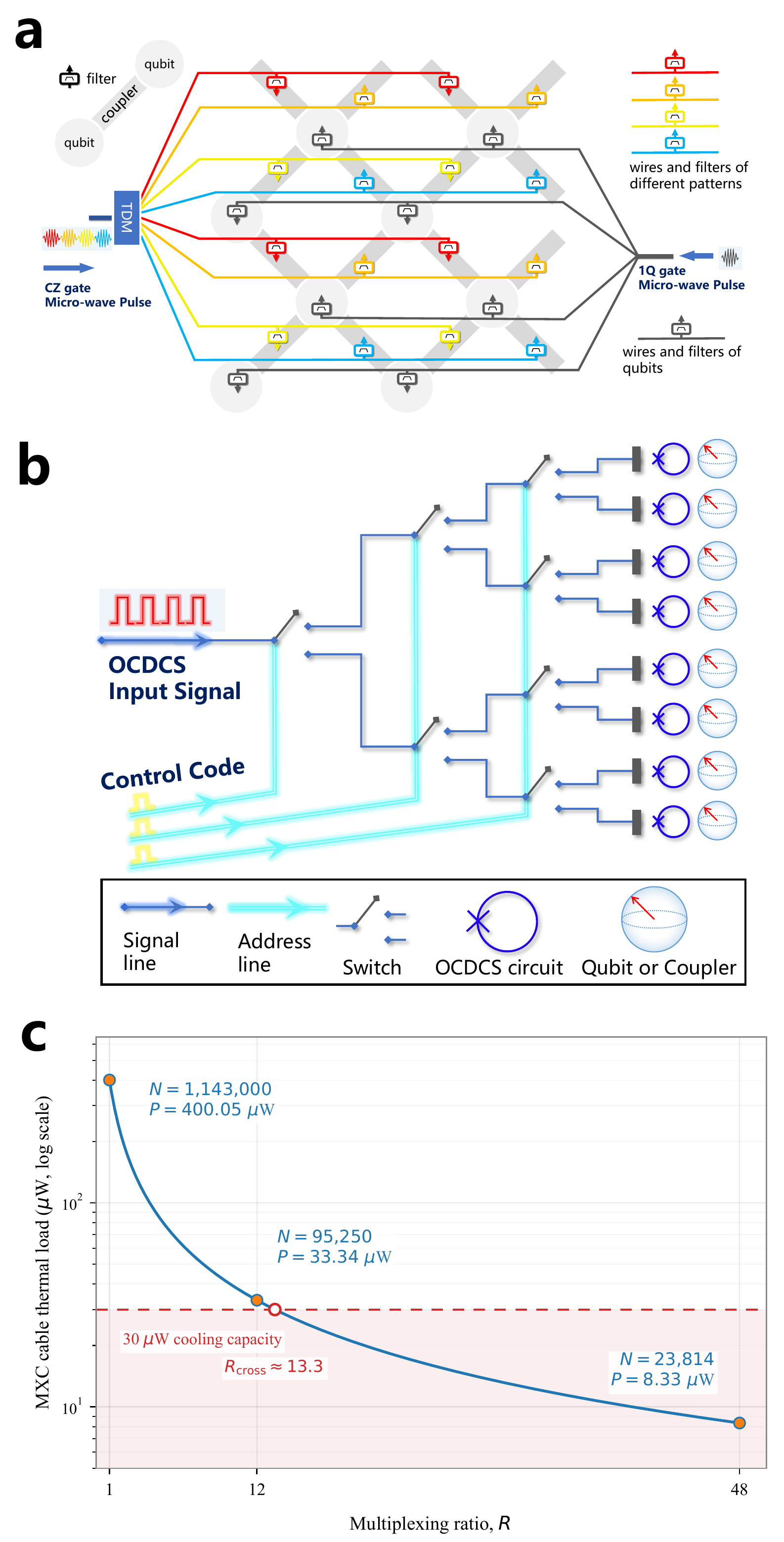}
    \caption{
    (a) The schematic diagram of control-line multiplexing qubit system. This diagram shows a 2d-periodic scalable structure. The colors of the wires and filters of the couplers mark the patterns of the couplers. The patterns are divided according to the surface code~\cite{fowler2012surface}. Only couplers in the same pattern need to perform parallel CZ gates. 
    (b) The OCDCS system with switch array~\cite{jiang2025chip}. 
    (c) Estimated passive MXC cable thermal load versus the control-line multiplexing ratio \(R\). Assuming \(381{,}000\) physical qubits, three dedicated control lines per qubit and the public NbTi FPC reference value of \(350~\mathrm{pW/channel}\)~\cite{DelftCircuits2026Brochure}, the solid curve follows \(P_{\mathrm{MXC}}=400.05/R~\mu\mathrm{W}\). Markers show \(R=1\), \(12\) and \(48\); the shaded region denotes loads below \(30~\mu\mathrm{W}\), with a crossover at \(R\approx13.3\). Other cryogenic heat loads are excluded.
    }
    \label{fig:multiplex_scheme}
\end{figure}

\begin{figure*}[hbtp]
    \centering 
    \includegraphics[width=\textwidth]{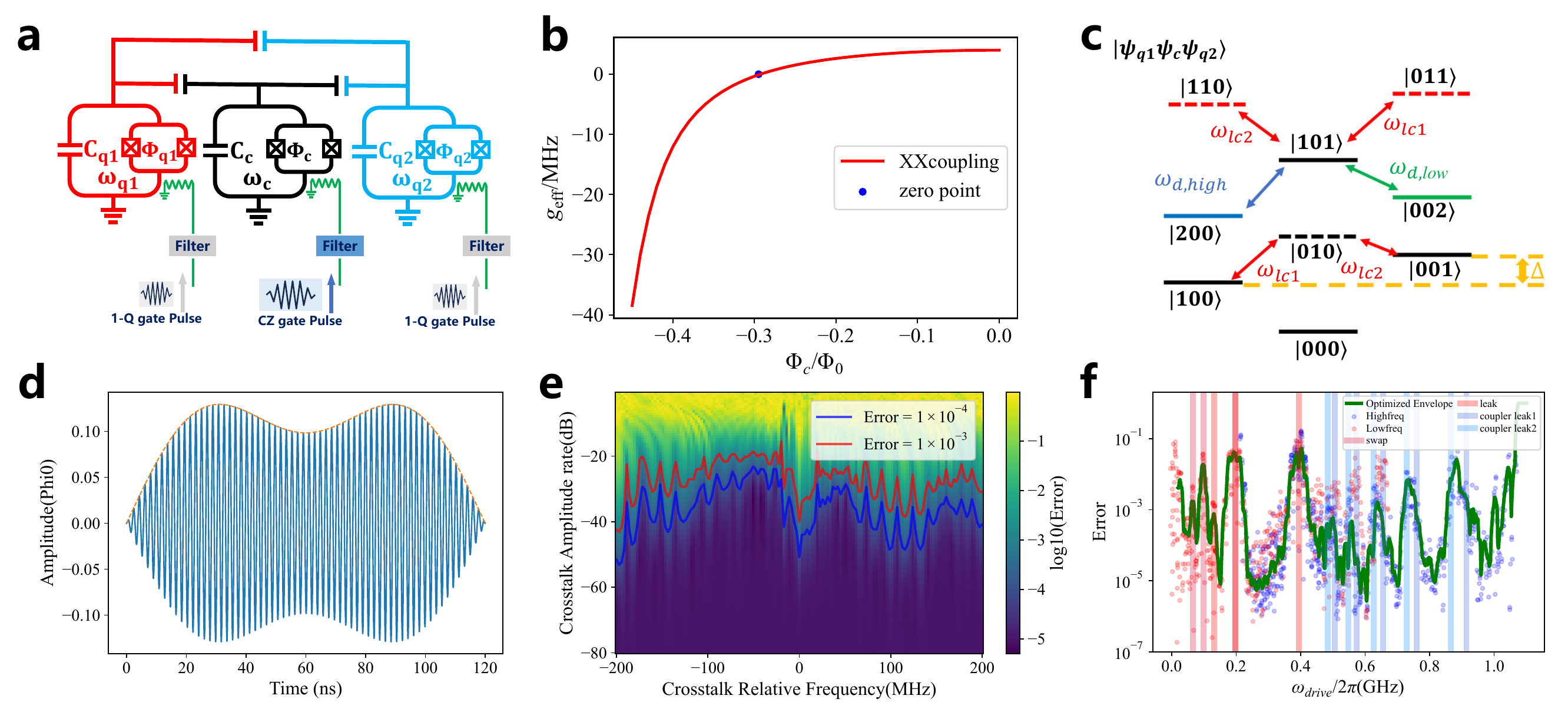}
    \caption{
    (a) Equivalent circuit of the qubit-coupler-qubit structure in \hyperref[fig:multiplex_scheme]{FDM diagram}. Coupling strength is tuned via the flux bias \(\Phi_c\)\cite{yan2018tunable}. 
    (b) Equivalent coupling strength vs. coupler flux bias. Blue circles indicate the coupling zero point. 
    (c) MP-CZ gate schematic. \(\omega_{d,high}\) and \(\omega_{d,low}\) denote the driving frequencies of high- and low-frequency MP-CZ gates, respectively. 
    (d) Driving waveform applied to the coupler flux bias. 
    (e) With MP-CZ gate frequency as zero, x-axis: crosstalk waveform frequency; y-axis: crosstalk-to-MP-CZ amplitude ratio. 
    The color represents the mixed signal error rate, including the original gate error (\(\approx 1\times 10^{-5}\)) and the crosstalk-induced error. 
    (f) Fidelity dependence on MP-CZ driving frequency. 
    \(Lowfreq\) and \(Highfreq\) denote low- and high-frequency MP-CZ gates, respectively. 
    The lower error rate from \{Lowfreq, Highfreq\} is selected for each point, smoothed to yield the optimized envelope.
    The vertical lines indicate predicted high-error points.  
    LC1 and LC2 denote leakage from q1 and q2 to the coupler, respectively. 
    }
    \label{fig:ACCZ_design}
\end{figure*}

\section{\label{section4_scalable} Scalable Scheme for Quantum Control}

We first describe the scalable control architecture underlying the multiplexing scheme. As illustrated in \autoref{fig:multiplex_scheme}~(a), multiple gate pulses are combined into a polychromatic microwave waveform and spectrally routed to the target qubits or couplers by band-pass filters. Frequency thus becomes an addressable control dimension, so the shared resource is a frequency-resolved control bus rather than a single physical line.
The architecture is also compatible with TDM~\cite{acharya2023multiplexed} and OCDCS~\cite{jiang2025chip, castellano2006static, deng2015working, bunyk2014architectural}. Because FDM requires a precise allocation of qubit and coupler frequencies, OCDCS is used to provide the bias-tuning capability needed to set and maintain the required idle frequencies. 

In error-correction cycles and many quantum algorithms, two-qubit gates are organized into parallel layers, in which only a subset of couplers is activated at a given time~\cite{fowler2012surface, arute2019quantum}. We exploit this structure by assigning simultaneously active couplers to the same control pattern. Within each pattern, different couplers are addressed by distinct CZ drive frequencies, enabling parallel operation while suppressing frequency-domain crosstalk. Couplers belonging to different patterns are not activated simultaneously and can therefore reuse the same spectral resources through TDM.
As an example, the periodic frequency arrangement analyzed in Supplementary Section~I.A combines 1:4 FDM for couplers, 1:4 TDM across coupler patterns, and 1:8 FDM for qubits, corresponding to an overall control-line reduction to 1/12 of a dedicated-line implementation under the gate parameters used in the current experiment. By contrast, under the idealized assumptions detailed in Supplementary Section~I.B, the same 32-qubit/64-coupler periodic unit can achieve 1:48. Larger allowed qubit-frequency differences, \(\Delta\), or a smaller minimum spacing between FDM tones can increase the usable frequency margin and support higher multiplexing ratios. 
\autoref{fig:multiplex_scheme}~(c) translates these multiplexing ratios into the corresponding passive MXC cable thermal load. Using the public NbTi FPC reference value of \(350~\mathrm{pW/channel}\)~\cite{DelftCircuits2026Brochure}, the estimate follows \(P_{\mathrm{MXC}}(R)=400.05/R~\mu\mathrm{W}\): the dedicated, \(1{:}12\) and \(1{:}48\) cases correspond to approximately \(400.05\), \(33.34\) and \(8.33~\mu\mathrm{W}\), respectively. The representative \(30~\mu\mathrm{W}\) cooling level is crossed at \(R\approx13.3\).


\section{\label{section2}Theoretical Analysis and Simulation for MP-CZ}
To support this architecture, we next analyze the MP-CZ gate mechanism and identify the frequency constraints that govern multiplexed operation.

\subsection{\label{subsection_theo} Theoretical Analysis}

The qubit-coupler-qubit structure shown in \autoref{fig:ACCZ_design}~(a) features tunable coupling strength \cite{yan2018tunable}. 
In an idle state, the coupling is turned off. 
\autoref{fig:ACCZ_design}~(b) shows the effective coupling strength vs. coupler flux phase. Simulation parameters are listed in \autoref{simu_paras}. 

We use the modulated coupling strength method for the microwave pulse CZ (MP-CZ) gate via the coupler Z-channel \cite{li2022realization, ma2025analysis}. 
The schematic diagram of the MP-CZ gate is shown in \autoref{fig:ACCZ_design}(c). 
Implementing an MP-CZ gate requires a \(\ket{101} \leftrightarrow \ket{200}\) swap or a \(\ket{101} \leftrightarrow \ket{002}\) swap. Thus, \(\omega_d\) is set to either \(\omega_2 - \omega_1 + |\eta_1|\) or \(\bigl|\omega_2 - \omega_1 - |\eta_2|\bigr|\); see Supplementary Section~II.A for the derivation. 

Define \(\Delta = \omega_{q2} - \omega_{q1} > 0\). The two branches have drive frequencies \(\omega_{d,\mathrm{high}}=|\omega_{101}-\omega_{200}|\approx\Delta+|\eta_1|\) and \(\omega_{d,\mathrm{low}}=|\omega_{101}-\omega_{002}|\approx\bigl|\Delta-|\eta_2|\bigr|\).
The driving type is selected based on specific requirements. 
The MP-CZ gate error depends strongly on \(\Delta\) \cite{ma2025analysis}. 
See Supplementary Section~II.B for details. 
When \(\Delta \approx |\eta|\), both types suffer from  \(\ket{2}\) state leakage. 
These error sources constrain qubit frequency arrangement. \(\Delta\) must avoid certain values and remain within bounds. 
For \(\Delta\approx (1+1/(n-1))|\eta|, n\in \{2,3,4,...\}\), low-frequency MP-CZ satisfies \(n\omega_{d,low} \approx \Delta \), causing swap error. 
For \(\Delta\approx (1+2/(n-1))|\eta|, n\in \{2,3,4,...\}\), low-frequency MP-CZ satisfies \(n\omega_{d,low} \approx \Delta+|\eta| \approx \omega_{d,high}\), causing leakage error. 
For \(n\omega_{d} \approx \overline{\omega_c}-\overline{\omega}_{q1/q2},\ n\in\{1,2,3,4,...\}\), the coupler might be excited. 
However, a considerable feasible solution space remains. The details are shown in Supplementary Figure~S1(b,c). 

\begin{table}[htbp]
    \centering
    \caption{Parameters of the qubits and coupler used in the simulation}
    \begin{tabular}{cccc}
        \toprule
        Parameter & Qubit 1 & Qubit 2 & Coupler \\
        \midrule
        \(f_{01,\max}(GHz)\) & 5.12 & 5.45 & 12.99 \\
        flux bias(\(\Phi_0\)) & -0.0 & -0.0 & -0.293 \\
        \(f_{01}(GHz)\) & 5.12 & 5.45 & 10.13 \\
        \(-|\eta|(GHz)\) & -0.198 & -0.196 & -0.144 \\
        \bottomrule
    \end{tabular}
    \label{simu_paras}
\end{table}

\begin{figure*}[htbp]
    \centering 
    \includegraphics[width=\textwidth]{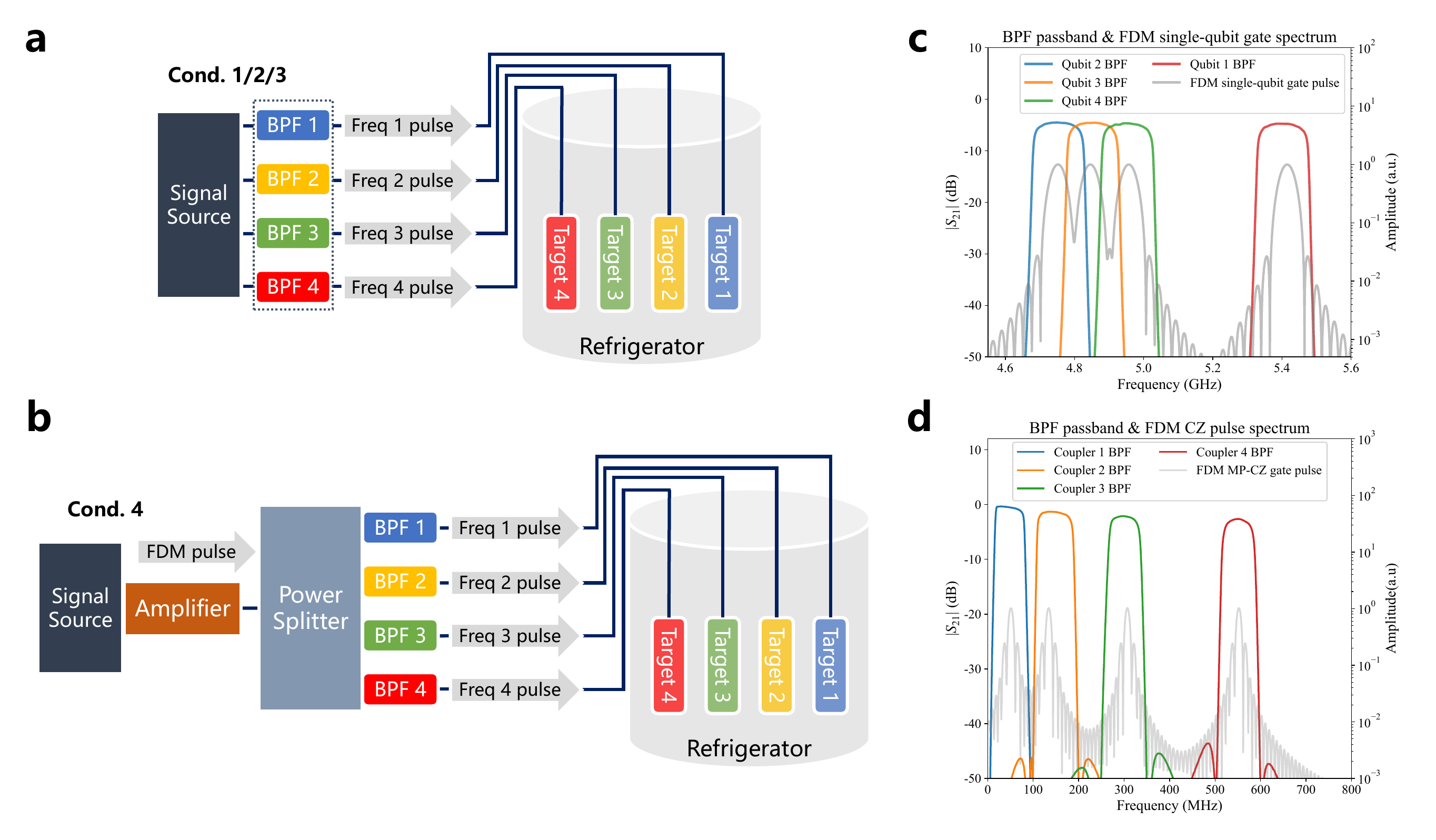}
    \caption{
    (a) Experimental system for Cond.~1/2/3. Cond.~1 has no BPF in the dotted box area. 
    (b) Experimental system for Cond.~4, the final configuration. 
    Both MP-CZ and single-qubit gate experiments follow this configuration. 
    For single-qubit experiments, the signal source is a 5.1~GHz microwave source and a 2~GSa/s AWG sideband; ``target'' denotes ``qubit''. 
    For MP-CZ experiments, the signal source is a 2~GSa/s AWG; ``target'' denotes ``coupler''. 
    (c) Filter passband and FDM pulse spectrum for FDM single-qubit gates. Right y-axis: FDM single-qubit pulse spectrum; left y-axis: four band-pass filter passbands. BPF denotes band-pass filter. 
    (d) Filter passbands and FDM pulse spectrum for FDM CZ gates. Right y-axis: FDM CZ pulse spectrum; left y-axis: four band-pass filter passbands. BPF denotes band-pass filter. 
    }
    \label{fig:exp_setup}
\end{figure*}


\begin{figure*}[htbp]
    \centering 
    \includegraphics[width=\textwidth]{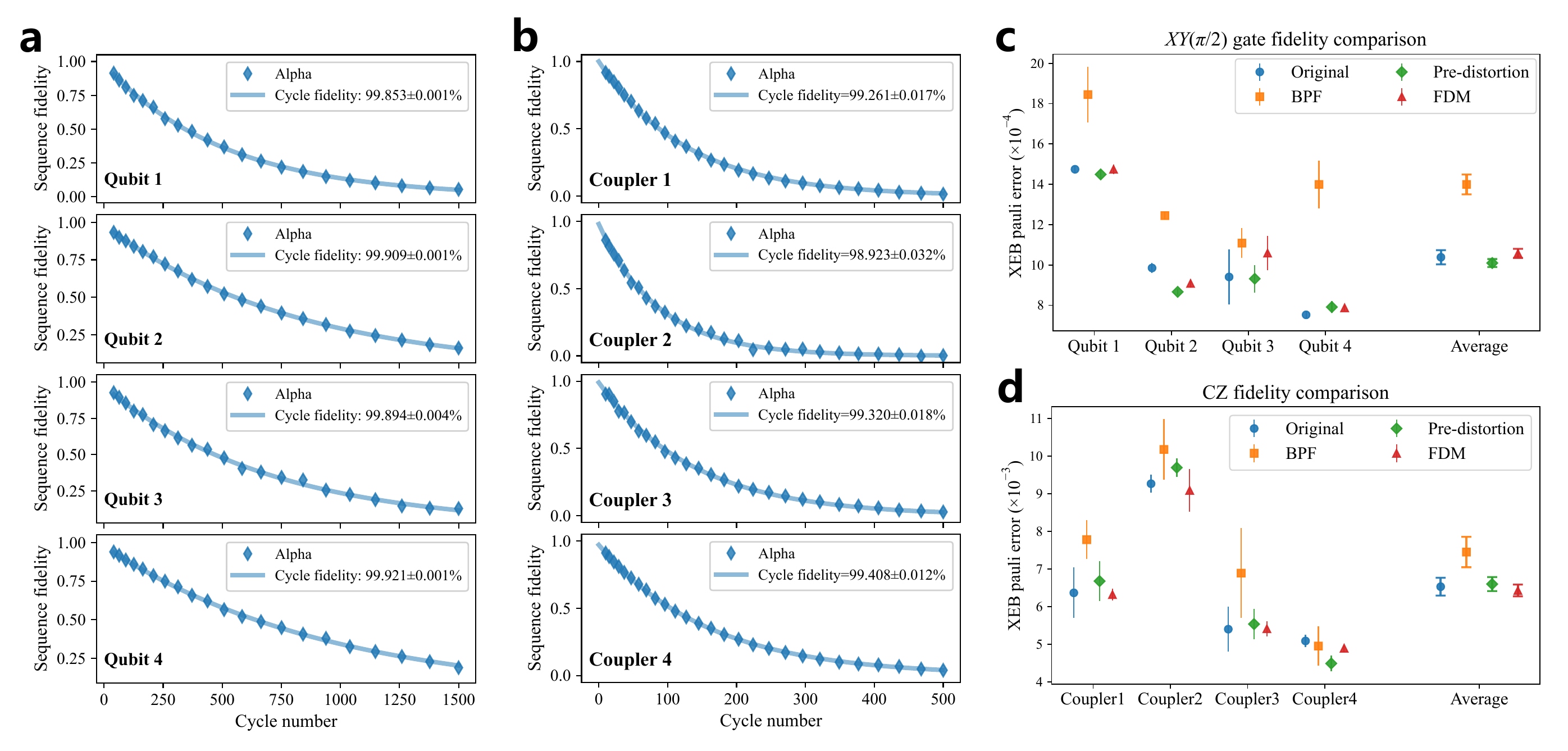}
    \caption{
    (a) XEB results of FDM single-qubit gates. 
    The per-cycle XEB fidelities of Qubits 1–4 are 99.853$\pm$0.001\%, 99.909$\pm$0.001\%, 99.894$\pm$0.004\% and 99.921$\pm$0.001\%. The average single-qubit gate fidelity is 99.894$\pm$0.002\%.
    (b) XEB results of the FDM MP-CZ gates. 
    The per-cycle XEB fidelities of Couplers 1–4 are 99.26$\pm$0.02\%, 98.92$\pm$0.03\%, 99.32$\pm$0.02\% and 99.41$\pm$0.01\%. MP-CZ fidelities are corrected by the product of the two participating $XY(\pi/2)$ fidelities in Supplementary Table~S1; the coupler--qubit mapping is given in Supplementary Section~VII. The MP-CZ fidelities of Couplers 1–4 are 99.37$\pm$0.02\%, 99.06$\pm$0.03\%, 99.48$\pm$0.02\% and 99.51$\pm$0.01\%. The average CZ fidelity is 99.35$\pm$0.01\%.
    (c) shows the statistical results for the \(XY(\pi/2)\) error under different conditions and qubits. 
    The 'Average' data are averaged over the four qubits. 
    Error bars denote 95\% confidence intervals, reflecting the fluctuations of the system and the fitting error. 
    (d) shows the statistical results for the MP-CZ error at different stages and different couplers. 
    The 'Average' data are averaged over the four couplers. 
    Error bars denote 95\% confidence intervals, reflecting the fluctuations of the system and the fitting error. 
    }
    \label{fig:exp_result}
\end{figure*}

\subsection{\label{subsection_simu} Simulation of the MP-CZ gate}
For convenience, we perform the simulation without considering the filters, which makes the driving amplitude represent the real flux amplitude. 
In our simulation, the decoherence error is not considered. 

Using the qubit and coupler parameters listed in \autoref{simu_paras} and the envelope shown in \autoref{fig:ACCZ_design}~(d) with a gate length of 120 ns, 
sweeping the frequency difference between the qubits, we simulate the evolution of the system and calculate the optimized infidelity of the gate. 
\autoref{fig:ACCZ_design}~(f) shows the dependency of optimized gate error on driving frequency. 
The vertical lines in \autoref{fig:ACCZ_design}~(f) represent the high-infidelity points predicted by the equations in \hyperref[subsection_theo]{Theoretical Analysis}. 
We can see that the high-error points are mostly in line with the predictions. 

\subsection{\label{subsection_crosstalk} Crosstalk-Induced Infidelity}
Using the parameters shown in \autoref{simu_paras}, we obtain an optimized MP-CZ with a simulated error below \(1\times 10^{-5}\). 
We apply an extra pulse with the same envelope and duration, but different amplitude and frequency to the coupler Z. We denote the extra pulse as the crosstalk pulse. 
Sweeping the frequency and amplitude of the crosstalk pulse, we obtain the result shown in \autoref{fig:ACCZ_design}~(e). 
In the FDM scheme, the crosstalk pulse is mainly caused by signals from other frequency bands that the filter has not rejected cleanly. 
Assuming the frequency spacing for frequency division multiplexing is 100~MHz, the filter should provide at least 50~dB of isolation at \(\pm 100\)~MHz to suppress the crosstalk-induced infidelity by around \(1\times 10^{-4}\). 


\section{\label{section3} Experimental demonstration}
We then turn to the experimental validation of the architecture, with particular focus on the parallel frequency-multiplexed single-qubit gates and the first demonstrated frequency-multiplexed two-qubit gates in a superconducting quantum processor.
\subsection{\label{subsection_expsys} Experimental System}
The experiment is conducted in a superconducting transmon qubit system with tunable couplers. 
Four qubits are used for single-qubit gate FDM experiments and four couplers for MP-CZ FDM experiments. 
The details of the device selection for both experiments are provided in Supplementary Section~VII.

To assess the effects of filtering, waveform compensation and shared-channel multitone control on gate performance, we compare four experimental conditions: 
(Cond. 1) Original — each target is controlled by a dedicated room-temperature electronics channel; 
(Cond. 2) BPF — band-pass filters are inserted into each signal path, followed by standard optimization and calibration; 
(Cond. 3) Pre-distortion — pre-distortion is applied to compensate for the BPF response; and 
(Cond. 4) FDM — a power splitter is added to enable four-target control through a single multitone channel. 
The BPFs are placed at room temperature to allow rapid reconfiguration among these conditions while keeping the cryogenic wiring and quantum processor unchanged. For this gate-level comparison, the relevant factor is the stable, calibratable frequency-dependent transfer function of the control path, rather than the physical location of the filters, allowing us to isolate the effects of filtering and frequency multiplexing.
The diagram of Cond.~1/2/3 is shown in \autoref{fig:exp_setup}~(a), and Cond.~4 in \autoref{fig:exp_setup}~(b). 
In Cond.~4, room-temperature electronics generate the control signals, which are split by a room-temperature power splitter and band-pass filters (BPF) into multiple lines at distinct frequencies. These signals are then delivered to the millikelvin-temperature superconducting chip. 

\subsection{\label{subsection_fdm1q} FDM Single-qubit Gates}

We employ a multi-frequency control signal to implement four simultaneous $XY(\pi/2)$ gates on distinct qubits. 
The experimental diagram is shown in \autoref{fig:exp_setup}, and the frequency spectrum of the composite control signal with the corresponding band-pass filter (BPF) transmission profiles is presented in \autoref{fig:exp_setup}~(c). 
Despite spectral overlap between adjacent qubits and finite BPF isolation, the following results show that crosstalk has negligible impact on gate fidelity after pre-distortion. 
With the room-temperature BPF, four parallel single-qubit gates are performed using a single 5.1~GHz microwave source and a 2~GSa/s AWG for sideband generation.

To benchmark the fidelity of the \(XY(\pi/2)\) gates, we perform cross-entropy benchmarking (XEB)~\cite{arute2019quantum}. As shown in \autoref{fig:exp_result}~(a), the average fidelity across the four qubits is \(99.894\pm0.002\%\). Panel (c) shows the XEB Pauli error for each qubit under four experimental conditions, with each data point averaged over five repetitions; the corresponding averages across all four qubits are also shown as 'Average'. Error bars denote 95\% confidence intervals. 

The BPFs initially cause a substantial increase in XEB Pauli error, whereas waveform pre-distortion largely compensates for their response, reducing the error to $(10.10\pm0.20)\times10^{-4}$, statistically indistinguishable from the no-BPF baseline of $(10.38\pm0.35)\times10^{-4}$. After adding the power splitter required to realize the full multiplexing configuration, the error remains $(10.57\pm0.23)\times10^{-4}$, again with no statistically significant degradation relative to the baseline. Thus, within the experimental resolution, the implemented FDM control chain introduces no measurable additional gate error after waveform pre-distortion.

\subsection{\label{subsection_fdmCZ} FDM MP-CZ Gates}
We use a multi-frequency signal to perform the four CZ gates. The spectrum of the signal and the BPFs we use are shown in \autoref{fig:exp_setup}~(d). 
To benchmark the fidelity of the four MP-CZ gates, we use cross-entropy benchmarking (XEB) \cite{arute2019quantum}. 
One of the FDM XEB results is shown in \autoref{fig:exp_result}~(b), in which the average CZ fidelity of the four couplers is \(99.35\pm0.01\)\%.

For all four conditions, the pulse duration is kept at 120 ns, with 4 ns of idle time before and after, totaling 128 ns. The \(XY(\pi/2)\) gate length is 40 ns.

The XEB Pauli error for each coupler under the four conditions is shown in \autoref{fig:exp_result}~(d). Each point is averaged over five XEB data sets measured at different times. \autoref{fig:exp_result}~(b) is one of the five XEB results obtained under the FDM condition. 
According to the data shown in \autoref{fig:exp_result}~(d), band-pass filters increase the XEB error rate due to pulse distortion. 
After pre-distortion, the error is suppressed and the overall XEB error returns to a level comparable to the baseline without BPFs. 
Comparing the average error of the original condition (\((6.53\pm0.24)\times10^{-3}\)) with that of the FDM condition (\((6.43\pm0.16)\times10^{-3}\)), FDM does not induce significant additional error. 

\section{\label{section5}Discussion}
Our results establish FDM gate control as a viable route to reducing the control-line overhead of superconducting quantum processors. The analysis of the MP-CZ gate identifies the frequency-allocation constraints and error mechanisms that govern parallel operation. Experimentally, we demonstrate 1:4 FDM of both single-qubit gates and MP-CZ gates without statistically significant multiplexing-induced degradation after waveform pre-distortion. These results show that frequency sharing can be introduced at the gate level without a measurable fidelity penalty, providing the missing gate-control element for an all-microwave architecture supporting universal quantum control. 

Extending this gate-level result to a scalable control architecture requires the spectral sharing provided by FDM to be coordinated with temporal resource sharing. FDM relies on frequency-selective routing of control tones, while OCDCS provides the idle-frequency tunability needed for spectral assignment~\cite{jiang2025chip}. Cryogenic switch arrays could add a complementary TDM layer by reallocating bias resources among qubit subsets and, when simultaneous operation is not required, sharing coupler-control resources. Related switch technologies have already been demonstrated in superconducting-control settings, including at the qubit stage~\cite{Hornibrook2015Architecture, Pechal2016Wallraff, 24Zotovatoswitch, zhao2026persistentcurrentswitch}. Together, FDM filtering, on-chip biasing and cryogenic switching form a candidate low-temperature control stack that could be co-integrated through through-silicon-via (TSV)/redistribution layer (RDL)-based packaging.  

At fault-tolerant scale, the periodically extensible scheme provides a quantitative projection of the required control-line and thermal resources. Using the estimates in Supplementary Section~I.C, for an RSA-2048-scale processor, the corresponding idealized conductive load is \(400.05~\mu\mathrm{W}\) with dedicated lines but \(8.33~\mu\mathrm{W}\) with multiplexing. Multiplexing brings the control-line contribution within the cited \(5\text{--}30~\mu\mathrm{W}\) MXC cooling range~\cite{tian2025high, krinner2019engineering, raicu2025cryogenic}. Realizing the projected 1:48 multiplexing ratio will require robust frequency allocation, mitigation of device variability, improved qubit coherence, integration of band-pass filters and OCDCS, and cryogenic switch arrays. Together, these results establish frequency-multiplexed control as a viable scaling architecture, bringing otherwise prohibitive control-line requirements within the capacity of existing dilution refrigerators and providing a critical step toward application-relevant fault-tolerant systems. 

\section*{Acknowledgments}
\begin{acknowledgments}
We thank the University of Science and Technology of China (USTC) Center for Micro- and Nanoscale Research and Fabrication for supporting the
sample fabrication and QuantumCTek Co., Ltd. for the manufacture
and maintenance of room-temperature electronics equipment.
This research was supported by the Quantum Science and
Technology-National Science and Technology Major Project (grant
numbers 2021ZD0300200 and 2023ZD0300200), special funds from Jinan Science and Technology Bureau and Jinan High Tech Zone Management
Committee, National Natural Science Foundation of China (grant numbers 92476203, 92476001, 12175003 and 12361161602), NSAF (grant number U2330201), Shandong Provincial Natural Science Foundation (grant number
ZR2022LLZ008), Cultivation Project of Shanghai Research Center for
Quantum Sciences (grant number LZPY2024), Key-Area Research and
the Innovation Program for Quantum Science and Technology (Grant No.
2024ZD0302600). M.G. was sponsored by National Natural Science Foundation of China (grant numbers T2322024 and 12474495), Shanghai Rising-Star Program (grant number 23QA1410000) and the Youth Innovation
Promotion Association of CAS (grant number 2022460). X. Zhu
acknowledges support from the New Cornerstone Science Foundation
through the Xplorer Prize and the Taishan Scholars Program.
\end{acknowledgments}


\section*{Competing interests} 
The authors declare no competing interests.

\bibliography{mybib}

\section*{Methods}

\subsection{Quantum processor and control setup}

All experiments were performed on a superconducting transmon processor with tunable
couplers. Four couplers were used for MP-CZ FDM and four qubits for \(XY(\pi/2)\) FDM;
the corresponding device selections are described in Supplementary Section~VII. \(XY(\pi/2)\) control signals were generated with a microwave source and a 2~GSa/s arbitrary waveform generator (AWG) for sideband generation. MP-CZ waveforms were generated with a 2~GSa/s AWG and applied to the coupler Z channels. The dedicated-channel, BPF-only,
pre-distorted and FDM configurations were individually optimized and calibrated before
comparison.

\subsection{Frequency assignment and multiplexing protocol}

Frequencies are reported either as ordinary frequencies \(f\) or as angular
frequencies \(\omega=2\pi f\). Each tone was assigned according to the qubit
\(f_{01}\) frequency or the required MP-CZ coupler-drive frequency. The signed
anharmonicities \(\eta_i<0\) follow the convention used in Supplementary Table~S1,
and \(|\eta_i|\) denotes their magnitude. For \(\Delta=\omega_{q2}-\omega_{q1}>0\), the two MP-CZ branches were \(\omega_{d,\mathrm{high}}\approx\Delta+|\eta_1|=\Delta-\eta_1\) and \(\omega_{d,\mathrm{low}}\approx|\omega_{101}-\omega_{002}|\simeq\Delta-|\eta_2|\) for the usual regime \(\Delta>|\eta_2|\).

The frequency arrangement based on the experimental gate parameters and its periodic
extension are described in Supplementary Section~I.A.
The idealized frequency plan assumed an approximately 1~GHz qubit-frequency window,
\(|\eta|/2\pi=210~\mathrm{MHz}\) and a 30~MHz tone-spacing criterion. The arrangement
based on the experimental gate parameters used gate durations of 40~ns for \(XY(\pi/2)\)
and 120~ns for MP-CZ
(Supplementary Section~I.A), whereas
the idealized projection used 100~ns and 200~ns durations, respectively, together with Hanning
envelopes for both gate classes, equal target and crosstalk amplitudes, negligible
in-band distortion and a coherence time sufficiently long that decoherence was not the
dominant error source
(Supplementary Section~I.B). The idealized
BPF profiles were used as frequency-routing models rather than device specifications.
The 30~MHz spacing criterion was applied to tones assigned to the same shared FDM
resource and activated simultaneously.

\subsection{Filter characterization and waveform compensation}

The complex \(S_{21}\) response of the room-temperature BPFs was measured with a vector
network analyzer and fitted to obtain the in-band group delay and temporal distortion.
The delay was checked by scanning the relative timing of the XY and Z channels
(Supplementary Section~VI.A). For both \(XY(\pi/2)\) and MP-CZ FDM,
the measured response was inverted to construct a finite-impulse-response (FIR)
pre-distortion kernel for the corresponding control path. The kernel was convolved
with the input waveform, and the resulting waveform was recalibrated at the chip
before comparison with the dedicated configuration. The FIR construction is described
in Supplementary Section~VI.B. 

\subsection{MP-CZ gate implementation and calibration}

The MP-CZ gate was implemented by modulating the effective coupling of a
qubit--coupler--qubit system through the coupler Z channel~\cite{li2022realization,
ma2025analysis}. The drive addressed the
\(\ket{101}\leftrightarrow\ket{200}\) or
\(\ket{101}\leftrightarrow\ket{002}\) transition, for which
\(\omega_d=\Delta+|\eta_1|\) and
\(\omega_d=\Delta-|\eta_2|\), respectively. The experimental MP-CZ pulse
duration was 120~ns, with 4~ns of idle time on either side, and the
\(XY(\pi/2)\) pulse duration was 40~ns. The pulse envelopes used in the MP-CZ
experiments are shown in Supplementary
Section~I.A; the condition-dependent
envelope choices and MP-CZ device selection are described in Supplementary
Section~VII.A, with the corresponding parameter
values listed in Supplementary Table~S1.

\subsection{Numerical model and crosstalk evaluation}

All quantum dynamical simulations used a four-level local truncation for each mode,
retaining \(\ket{0}\) through \(\ket{3}\), together with the device parameters listed in
\autoref{simu_paras}. The three-level Hamiltonians used later in the analytical discussion
are projections of this four-level numerical model and are not the propagation model used
to generate the simulation data. Except for the idealized
frequency-assignment analysis in Supplementary Section~I.B,
the gate durations, waveform types and control parameters were those used in the
experiment. The frequency-dependent MP-CZ infidelity was evaluated while sweeping the
qubit-frequency difference, and the lower-error MP-CZ branch was selected at each
frequency point to form the optimized envelope. Decoherence and filter distortion were
omitted when isolating frequency-dependent contributions. Crosstalk was evaluated by
adding a pulse with the same envelope and duration while varying its relative frequency
and amplitude. In the idealized analysis, the device parameters in \autoref{simu_paras}
were retained, while the 100~ns \(XY(\pi/2)\) and 200~ns MP-CZ pulses had
Hanning envelopes and equal target and crosstalk amplitudes, with the additional assumptions
specified in Supplementary Section~I.B.
The Hamiltonian parameterization, numerical truncation and analytical
projection are given in Supplementary Section~II.A; the numerical input parameters are
listed in Supplementary Section~VIII. The initial error-source classification is given
in Supplementary Section~II.B.

\subsection{Gate benchmarking and statistical analysis}

Performance of the \(XY(\pi/2)\) and MP-CZ gates was characterized using cross-entropy benchmarking
(XEB)~\cite{arute2019quantum}. For each target and configuration, five XEB acquisitions
were performed at different times. The XEB Pauli error and per-cycle fidelity were
obtained by fitting each XEB data set; the reported averages combine the four selected
qubits or couplers. For the MP-CZ benchmark, one XEB cycle contains one MP-CZ gate and
two randomly selected \(XY(\pi/2)\) gates, one on each participating qubit. Thus, \(F_{\mathrm{XEB}}\) denotes
the per-cycle fidelity extracted from the exponential
sequence-fidelity fit. For coupler \(c_i=(q_{i,1},q_{i,2})\), with the mapping given in
Supplementary Section~VII.A, define \(F_{1q}^{(i)}=F_{XY(\pi/2),q_{i,1}}F_{XY(\pi/2),q_{i,2}}\),
where \(F_{XY(\pi/2),q}=1-\epsilon_{XY(\pi/2),q}\). The reported MP-CZ fidelity is
\(F_{\mathrm{MP-CZ}}^{(i)}=F_{\mathrm{XEB}}^{(i)}/F_{1q}^{(i)}\),
where the single-qubit calibration errors are listed in Supplementary Table~S1. We treat
\(F_{1q}^{(i)}\) as the independent single-qubit calibration factor for the two qubits
participating in \(c_i\), so the corrected value is a per-cycle MP-CZ fidelity estimate.
Error bars denote 95\% confidence intervals combining temporal fluctuations and fitting
uncertainty. The comparison between the
FDM and Origin configurations was based on their mean XEB Pauli errors. We defined
\[
\Delta\bar{\epsilon}=\bar{\epsilon}_{\mathrm{FDM}}
-\bar{\epsilon}_{\mathrm{Origin}},
\]
where ``Origin'' denotes the Original condition (Cond.~1 in the main text), and the
barred quantities denote the mean errors over the four selected qubits or couplers.
Let \(\sigma_{\mathrm{FDM}}\) and \(\sigma_{\mathrm{Origin}}\) denote the corresponding
standard uncertainties obtained from the repeated XEB fits. Under an independent
normal approximation, the probability that the mean FDM error exceeded the mean
Origin error was evaluated as
\[
P_{\mathrm{FDM>Origin}}=\Phi\!\left(
\frac{\Delta\bar{\epsilon}}
{\sqrt{\sigma_{\mathrm{FDM}}^2+\sigma_{\mathrm{Origin}}^2}}
\right),
\]
where \(\Phi\) is the standard-normal cumulative distribution function. A statistically
significant increase was declared only when
\(P_{\mathrm{FDM>Origin}}>0.95\); otherwise, the FDM error was reported as not
significantly higher than the Origin error. For the MP-CZ comparison, the resulting
probability was approximately 0.24.

\subsection{System-scale control-line and thermal-load projection}

The system-scale projection extended the periodic frequency arrangement to a
resource estimate of 381,000 physical qubits
\cite{mundada2026heterogeneous}. The dedicated baseline assumed three control lines per
qubit~\cite{gao2025establishing, he2025experimental, google2025quantum,
sivak2025reinforcement}, giving
\(N_{\mathrm{ded}}=3\times381{,}000=1{,}143{,}000\). For the idealized
32-qubit/64-coupler unit, the multiplexed line count was
\[
\begin{aligned}
N_{\mathrm{mux}}&=
\left\lceil\frac{381{,}000}{32}\right\rceil+
\left\lceil\frac{2\times381{,}000}{64}\right\rceil\\
&=23{,}814,
\end{aligned}
\]
where the first term counts qubit-FDM lines and the second counts coupler-branch lines;
the factor of two in the second term accounts for two couplers per qubit in the
periodic unit. The frequency-arrangement inputs for the experimentally parameterized
and idealized projections are
described in Supplementary Section~I.A and Section~I.B, respectively.

The MXC conductive load was calculated as
\[
P_{\mathrm{MXC}}=N_{\mathrm{line}}p_{\mathrm{line}},
\]
where \(N_{\mathrm{line}}\) and \(p_{\mathrm{line}}\) denote the line count and
per-line conductive load, respectively. We used \(p_{\mathrm{line}}=350~\mathrm{pW/channel}=0.350~\mathrm{nW/channel}\), the public NbTi FPC signal-line value reported for the 100~mK--10~mK segment~\cite{DelftCircuits2026Brochure}. This manufacturer-reported value serves as a reference estimate for the passive conductive contribution, rather than as a direct measurement of the present device. 
The resulting conductive loads were compared with the representative
\(5\text{--}30~\mu\mathrm{W}\) MXC cooling range~\cite{tian2025high,
krinner2019engineering, raicu2025cryogenic}. This estimate includes the conductive
load of the control lines only; parasitic loads from filters, switches, splitters,
packaging and on-chip sources are excluded. The 1:48 line-reduction factor used in
this projection was taken from the idealized frequency plan in Supplementary Section~I.B.

\section*{Data Availability}
All source data supporting the findings of this study are available from the
corresponding author upon reasonable request.

\section*{Code Availability}
The custom code used for the numerical simulations and data analysis is available
from the corresponding author upon reasonable request.

\clearpage
\appendix
\renewcommand{\thefigure}{S\arabic{figure}}  
\renewcommand{\theequation}{S\arabic{equation}}  
\renewcommand{\thetable}{S\arabic{table}}
\renewcommand{\theHfigure}{S\arabic{figure}}  
\renewcommand{\theHtable}{S\arabic{table}}
\setcounter{figure}{0}  
\setcounter{equation}{0}  
\setcounter{table}{0}

\newpage
\setcounter{page}{1}

\begin{widetext}
\begin{center}
    \textbf{\large Supplementary Material for ``A Control Line Multiplexing Framework for Scalable Quantum Computing''}
\end{center}
\medskip  
\end{widetext}

\phantomsection
\pdfbookmark[0]{Supplementary Material}{sec:supplementary-material}

\section{\label{appendix_scable_frequency_arrangement}I. Scalable Frequency Arrangement for Multiplexing}
\subsection{\label{appendix_scalable_frequency_arrangement} I.A Scalable Frequency Arrangement}
\begin{figure*}
    \centering
    \includegraphics[width=0.9\linewidth]{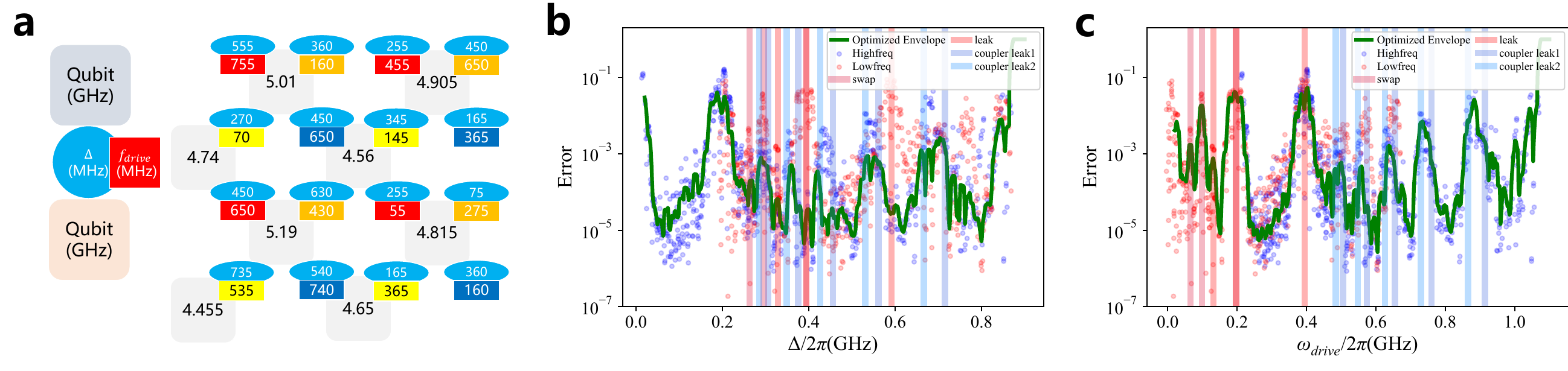}
    \caption{ 
    (a) Example frequency arrangement for the qubits. 
    Using the parameters shown in \autoref{simu_paras}, sweeping the frequency difference between the qubits, we simulate the evolution of the system and calculate the infidelity of the gate. (b, c) show the corresponding results. (b) shows the dependency of CZ fidelity on the frequency difference of the qubits, and (c) shows the dependency of fidelity on the MP-CZ driving frequency. 
    \(Lowfreq\) and \(Highfreq\) denote the low- and high-frequency MP-CZ gates, respectively. 
    We select the lower error rate from {Lowfreq, Highfreq} for each independent value, smooth the data and obtain the optimized envelope.
    The vertical lines represent high-error points predicted by the error model given in \hyperref[appendix_errorsour]{Supplementary Section~II.B}.  
    LC1 represents the leakage from q1 to the coupler, and LC2 represents the leakage from q2 to the coupler.
    }
    \label{fig:Scalable_Multiplexing}
\end{figure*}

\begin{figure*}[htbp]
    \centering
    \includegraphics[width=0.9\linewidth]{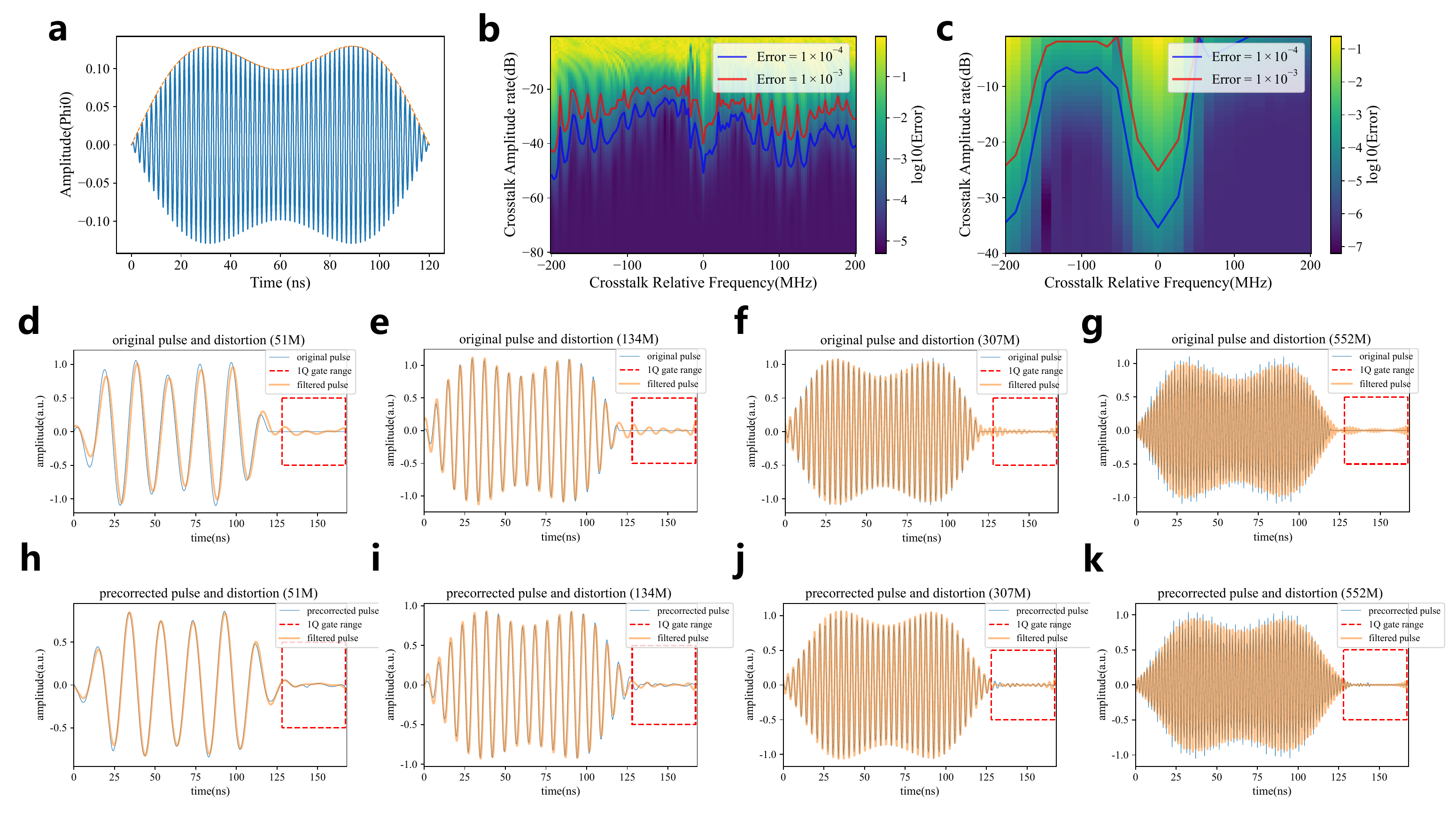}
    \caption{ 
    (a) MP-CZ pulse. (b) Crosstalk-induced infidelity of the 120~ns MP-CZ gate, including the original gate error (\(\approx 1\times10^{-5}\)). (c) Crosstalk-induced infidelity of the 40~ns $X_{\pi/2}$ gate using a Hanning envelope, including the original gate error (\(\approx 1\times10^{-7}\)).
    In (b) and (c), x: crosstalk frequency offset; y: \(20\log_{10}(A_{\mathrm{xtalk}}/A_{\mathrm{gate}})\) (dB); colour: \(\log_{10}\) of the error rate. 
    (d–g) show the distortion of pulses at 51, 134, 307 and 552 MHz after the BPF. 
    (h–k) show the pre-distorted pulses and the filtered waveforms at the same four frequencies. 
    (d) and (h) show the MP-CZ pulse for G0403.
    (e) and (i) show the MP-CZ pulse for G1612.
    (f) and (j) show the MP-CZ pulse for G0201.
    (g) and (k) show the MP-CZ pulse for G1413.
    }
    \label{fig:Harmonic_and_FDM}
\end{figure*}

According to \hyperref[appendix_errorsour]{Supplementary Section~II.B}, the high-frequency MP-CZ infidelity is mainly caused by the excitation of the coupler, and the low-frequency MP-CZ infidelity is caused by \(\ket{2}\) state leakage, swap error and the excitation of the coupler.

Since the driving frequency is determined by the frequency difference between the qubits, being \(\omega_{d,\mathrm{high}}\simeq\Delta+|\eta_1|\) for the high-frequency branch and \(\omega_{d,\mathrm{low}}\simeq\Delta-|\eta_2|\) for the low-frequency branch, we first need to analyze the gate infidelity as a function of the frequency difference between the qubits. For the following low-frequency resonance analysis, we assume \(\Delta>|\eta_2|\).
This yields a relation between the qubit frequency difference and the coupler excitation.
For high-frequency MP-CZ, when the frequency difference between the qubits is \(\Delta \approx \frac{\overline{\omega_c} - \overline{\omega_1}}{n} - |\eta_1|\) or \(\Delta \approx \frac{\overline{\omega_c} - \overline{\omega_1}}{n+1} - \frac{n}{n+1}|\eta_1|\), the coupler will be excited.
The details are given in \autoref{coupler_excitation_analysis}.
For low-frequency MP-CZ, the swap and leakage channels occur at different frequency differences,
\[
\begin{aligned}
\text{swap:}\quad &\Delta\simeq\left(1+\frac{1}{n-1}\right)|\eta_2|,\\
\text{leakage:}\quad &\Delta\simeq|\eta_2|+\frac{|\eta_1|+|\eta_2|}{n-1},\qquad n=2,3,\ldots
\end{aligned}
\]
where \(n\) denotes the drive harmonic order.

We sweep the frequency difference between the qubits and simulate the infidelity of the gate. The gate length is 120~ns, and the pulse is shown in \autoref{fig:Harmonic_and_FDM}~(a).
The results are shown in \autoref{fig:Scalable_Multiplexing}~(b); the error increases when \(\Delta\) satisfies \(\Delta \approx \frac{\overline{\omega_c} - \overline{\omega_1}}{n} - |\eta_1|\), \(\frac{\overline{\omega_c} - \overline{\omega_1}}{n+1} - \frac{n}{n+1}|\eta_1|\)
\(\Delta \approx \left(1+\frac{1}{n-1}\right)|\eta_2|\), or \(\Delta \approx |\eta_2|+\frac{|\eta_1|+|\eta_2|}{n-1}\). 

Considerable freedom nevertheless remains in choosing the frequencies. According to
\autoref{fig:Scalable_Multiplexing}(b), a high-fidelity MP-CZ can be performed when
\(\Delta/2\pi\) lies approximately in the intervals \(30\text{--}160\),
\(230\text{--}680\), or \(720\text{--}830~\mathrm{MHz}\). Accounting for
\autoref{fig:Scalable_Multiplexing}(c), the MP-CZ drive frequencies can be chosen near
\(50,150,250,350,450,550,650,750,850~\mathrm{MHz}\) for the present FDM plan.
For convenience, we consider \(|\eta|/2\pi=200~\mathrm{MHz}\), and try to arrange the frequencies of qubits. A feasible solution is shown in \autoref{fig:Scalable_Multiplexing}(a).

\subsection{\label{appendix_ideal_multiplexing_ratio} I.B Idealized Multiplexing Ratio Under the Assumed Conditions}
We estimate the multiplexing capacity of the present frequency plan under an idealized set
of assumptions. The coherence time is assumed to be sufficiently long that decoherence is
not the dominant error source for a 100~ns \(XY(\pi/2)\) gate or a 200~ns MP-CZ gate. The
target and crosstalk tones are taken to have equal amplitudes, and the band-pass filters
(BPFs) are assumed to introduce negligible in-band waveform distortion. For the frequency
assignment below, we take \(\lvert\eta\rvert/2\pi=210~\mathrm{MHz}\) and restrict the
qubit frequencies to an approximately \(1~\mathrm{GHz}\)-wide window, which defines the
available spectral span.

For a fixed normalized envelope, increasing the gate duration narrows the main spectral
lobe. The MP-CZ pulse is modeled with a Hanning envelope, as adopted for the \(XY(\pi/2)\)
gate. The combination of longer gate duration and Hanning-window sidelobe suppression
reduces spectral overlap and out-of-band components. Under the negligible-distortion BPF
assumption, this spectral concentration allows narrower passbands. The following analysis
uses these assumptions to assess FDM for both \(XY(\pi/2)\) and MP-CZ control.

\begin{figure*}[htbp]
    \centering
    \subfigure[\(XY(\pi/2)\) pulse spectrum and BPF response]{
        \includegraphics[height=0.17\linewidth,keepaspectratio]{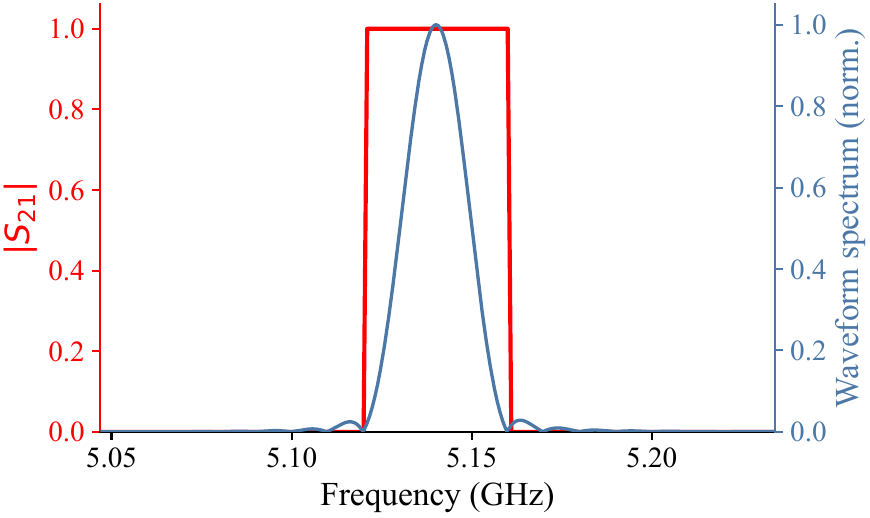}
        \label{fig:ideal_1q_filter}
    }
    \hfill
    \subfigure[100~ns \(XY(\pi/2)\) waveform]{
        \includegraphics[height=0.17\linewidth,keepaspectratio]{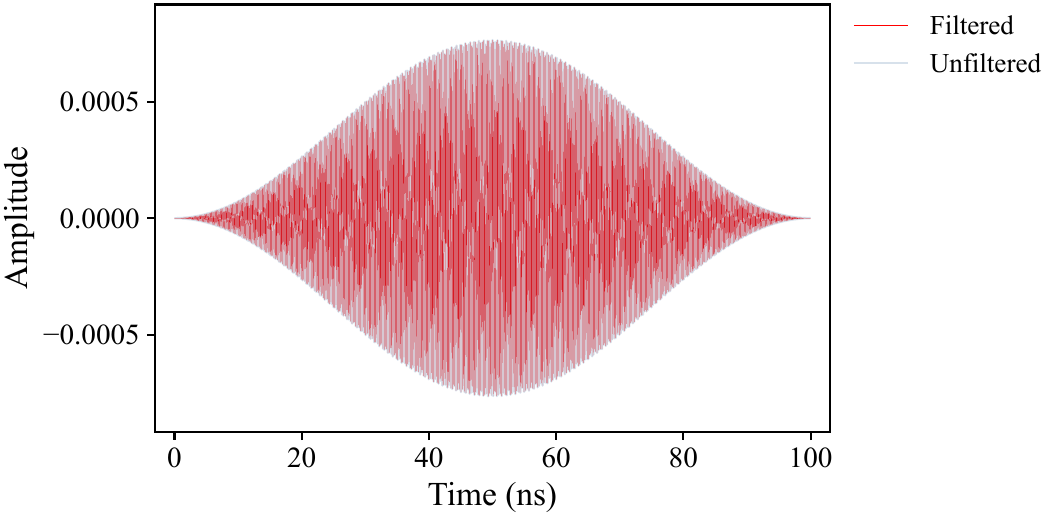}
        \label{fig:ideal_1q_waveform}
    }
    \hfill
    \subfigure[\(XY(\pi/2)\) crosstalk error]{
        \includegraphics[height=0.17\linewidth,keepaspectratio]{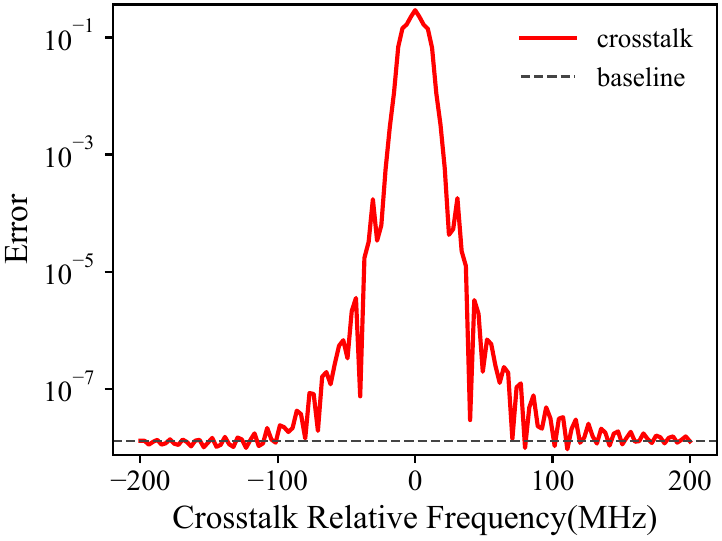}
        \label{fig:ideal_1q_crosstalk}
    }
    \caption{Idealized \(XY(\pi/2)\) FDM analysis. (a) BPF transmission and normalized
    waveform spectrum. (b) Filtered and unfiltered 100~ns \(XY(\pi/2)\) waveforms.
    (c) Gate error versus the frequency offset of an equal-amplitude crosstalk tone.
    These panels support the adopted frequency-spacing criterion and the negligible-distortion
    BPF assumption.}
    \label{fig:ideal_1q_evidence}
\end{figure*}
\begin{figure*}[htbp]
    \centering
    \subfigure[MP-CZ pulse spectrum and BPF response]{
        \includegraphics[height=0.17\linewidth,keepaspectratio]{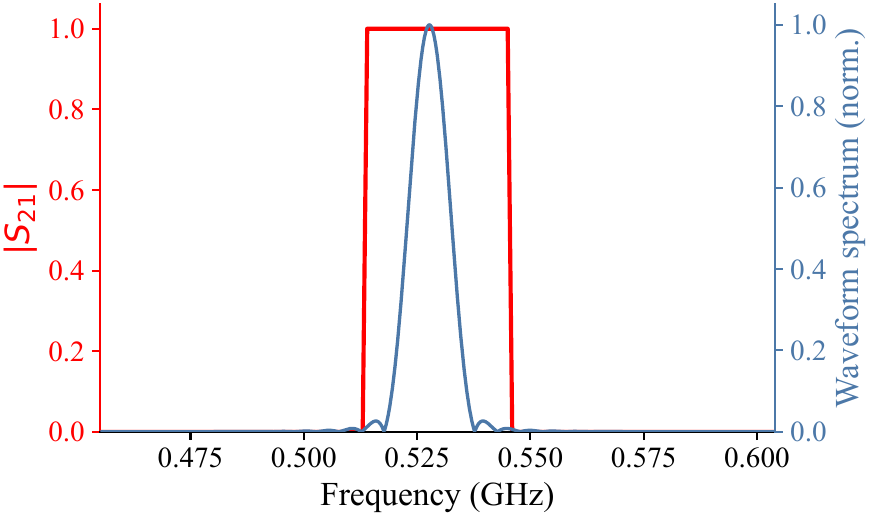}
        \label{fig:ideal_2q_filter}
    }
    \hfill
    \subfigure[200~ns MP-CZ waveform]{
        \includegraphics[height=0.17\linewidth,keepaspectratio]{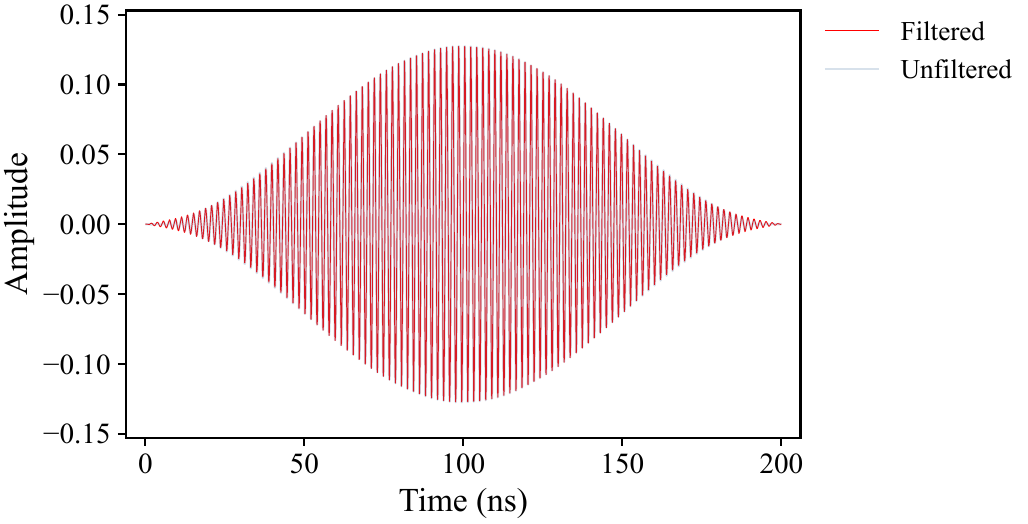}
        \label{fig:ideal_2q_waveform}
    }
    \hfill
    \subfigure[MP-CZ crosstalk error]{
        \includegraphics[height=0.17\linewidth,keepaspectratio]{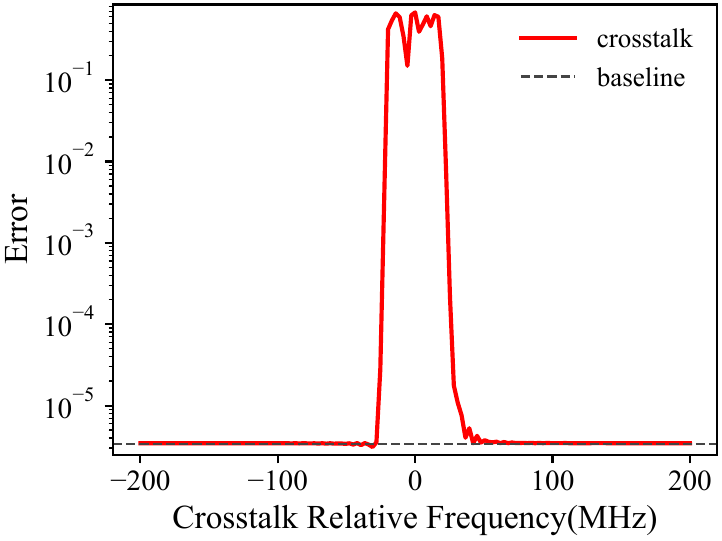}
        \label{fig:ideal_2q_crosstalk}
    }
    \caption{Idealized MP-CZ FDM analysis. (a) BPF transmission and normalized MP-CZ
    waveform spectrum. (b) Filtered and unfiltered 200~ns MP-CZ waveforms. (c) MP-CZ
    gate error versus the frequency offset of an equal-amplitude crosstalk tone. The
    crosstalk sweep provides the MP-CZ frequency-spacing criterion.}
    \label{fig:ideal_2q_evidence}
\end{figure*}

Figure~\ref{fig:ideal_2q_evidence}(a) shows that the MP-CZ spectrum lies within the BPF
passband and that out-of-band components are attenuated.
Figure~\ref{fig:ideal_2q_evidence}(b) shows that the 200~ns MP-CZ waveform is preserved
under the assumed filter response. The BPF model is used only for the present idealized
analysis; finite isolation, group delay, insertion loss, and calibration bandwidth are not
included.

Figure~\ref{fig:ideal_2q_evidence}(c) shows the corresponding MP-CZ crosstalk result.
The crosstalk error decreases sharply over the tens-of-megahertz transition region and is
already close to the baseline at an offset of approximately 30~MHz on the plotted positive-
offset side. We use 30~MHz as the nominal MP-CZ tone spacing in the present idealized
assignment. This engineering criterion is model-dependent and should not be interpreted as
a universal threshold for negligible MP-CZ error. Because the calculation uses one
equal-amplitude crosstalk tone and an idealized filter response, it is not a total
multi-tone, device-variation, and filter-calibration error budget.

The two crosstalk curves do not imply identical error thresholds at 30~MHz. Instead, the
same 30~MHz nominal spacing is adopted for both gate classes to define a common frequency
grid for tones assigned to the same FDM resource and activated simultaneously. This is a
design choice for the present idealized frequency plan, rather than a universal crosstalk
criterion.
Accordingly, the idealized BPF profiles in Figures~\ref{fig:ideal_1q_evidence}(a) and
\ref{fig:ideal_2q_evidence}(a) should be viewed as routing models rather than required
device specifications. Once the crosstalk-error criterion in Figures~\ref{fig:ideal_1q_evidence}(c)
and \ref{fig:ideal_2q_evidence}(c) is satisfied and in-band waveform distortion remains
negligible, the filter edge steepness can in principle be relaxed.

We next apply this common frequency grid to the frequency assignment. Let
\(\omega_{q1,i}\) and \(\omega_{q2,i}\) denote the transition angular frequencies of the
qubits participating in the \(i\)th MP-CZ gate, and define
\(\Delta_i=\omega_{q2,i}-\omega_{q1,i}>0\). The corresponding MP-CZ drive angular
frequencies are selected according to the high- and low-frequency branches,
\(\omega_{d,i}\simeq\Delta_i+\lvert\eta\rvert\) and
\(\omega_{d,i}\simeq\lvert\Delta_i-\lvert\eta\rvert\rvert\) for the high- and
low-frequency branches, respectively, where the low-frequency form retains the absolute
value when \(\Delta_i<\lvert\eta\rvert\). For each simultaneously active
FDM group, every pair of tones assigned to the same shared resource must satisfy
\(\lvert\omega_a-\omega_b\rvert/(2\pi)\geq\delta f_{\min}=30~\mathrm{MHz}\) and avoid
the leakage, swap, and coupler-excitation resonances derived above. Periodic assignments
then permit spectral-resource reuse across non-simultaneous TDM patterns.

\begin{figure*}[htbp]
    \centering
    \includegraphics[width=0.92\linewidth]{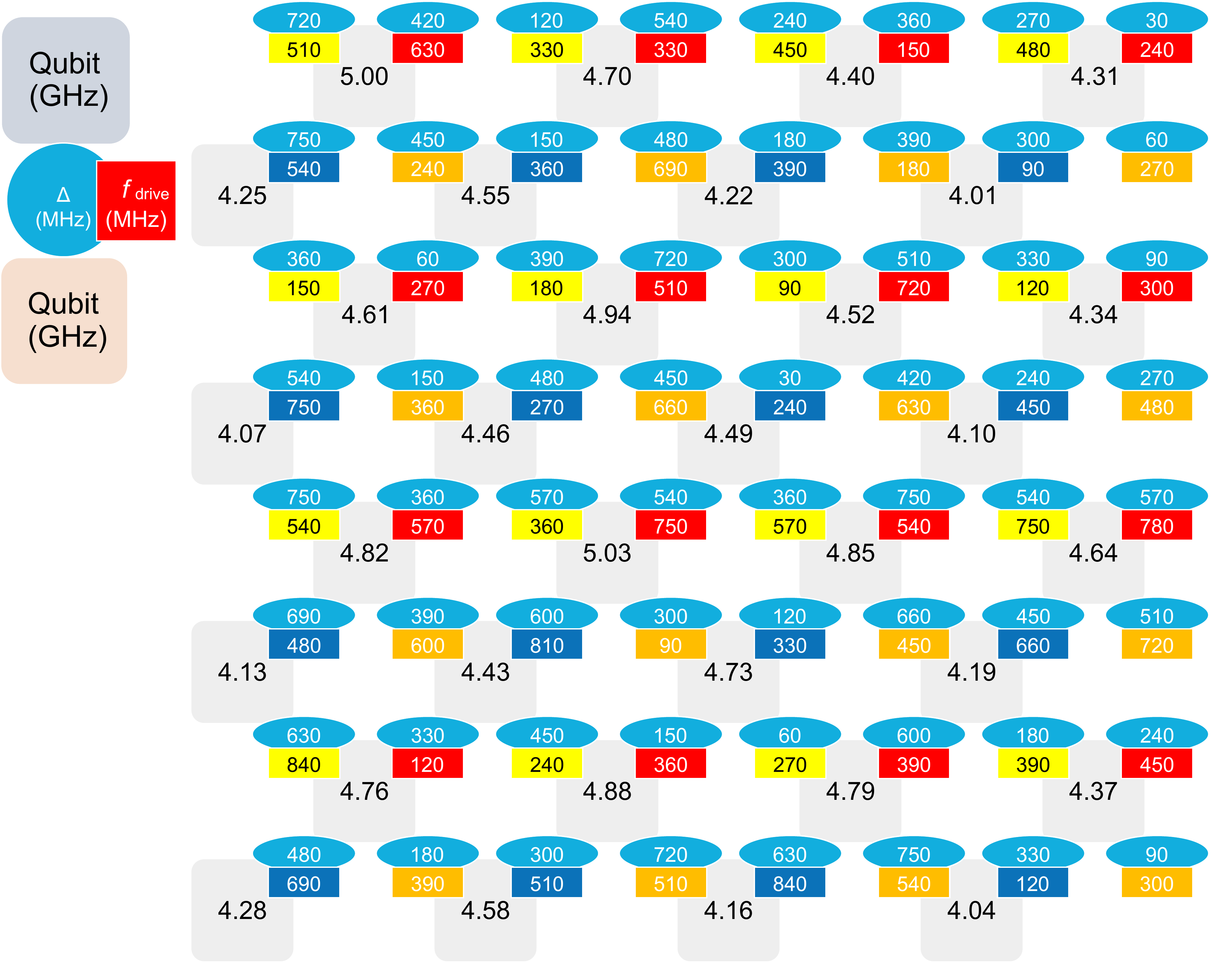}
    \caption{Idealized frequency assignment for \(\lvert\eta\rvert/2\pi=210~\mathrm{MHz}\).
    The diagram lists the qubit frequencies in GHz, the associated frequency differences
    \(\Delta/2\pi\), and the corresponding MP-CZ drive frequencies \(\omega_d/2\pi\), in MHz.
    Simultaneous tones are separated by at least 30~MHz and avoid the resonance exclusions. 
    This frequency assignment serves as a nominal design guide. 
    In practice, this nominal frequency assignment may require device-specific fine-tuning to accommodate additional constraints beyond those on the drive frequencies, such as near-resonances between the \(f_{01}\) and \(f_{12}\) transitions of different qubits.
    }
    \label{fig:ideal_frequency_assignment}
\end{figure*}

The resulting assignment (Figure~\ref{fig:ideal_frequency_assignment}) addresses 32 qubits
through one FDM line. The 64 couplers are partitioned into four non-overlapping TDM groups
activated in separate time slots, with each group addressed by one FDM branch line. The
coupler path therefore contains one 1:4 TDM switch and four FDM branch lines. At the
external interface, these branches are accessed through one switched path, yielding one
qubit FDM line and one switched coupler path, or two external control lines in total.

Relative to 32 dedicated qubit lines and 64 dedicated coupler lines, the idealized external
line count is reduced from
\begin{equation}
    N_{\mathrm{dedicated}}=32+64=96
\end{equation}
to
\begin{equation}
    N_{\mathrm{ideal}}=1_{\mathrm{qubit,FDM}}+1_{\mathrm{coupler,TDM/FDM}}=2.
\end{equation}
The corresponding reduction factor for this frequency plan is
\begin{equation}
    R_{\mathrm{line}}=\frac{N_{\mathrm{dedicated}}}{N_{\mathrm{ideal}}}=48.
\end{equation}
Thus, under the present assumptions, including the \(1.02~\mathrm{GHz}\)-wide qubit-frequency
window and this specific 32-qubit/64-coupler assignment, the same layout can achieve 1:48.
It is not a universal theoretical maximum: changing the gate durations, frequency-spacing
criterion, anharmonicity, scheduling pattern, or filter response changes the attainable ratio.
Internal fan-out, four FDM branches, the switch, splitters, bias sources, calibration
resources, and packaging connections remain necessary. Finite coherence time, filter
isolation and group delay, waveform distortion, frequency drift, switch and splitter loss,
calibration overhead, and additional DC-bias connections will further reduce the practical
multiplexing ratio.

\subsection{\label{appendix_rsa2048_mxc_thermal_load} I.C Idealized MXC Thermal-Load Estimate Using a Public NbTi FPC Reference}

We translate the idealized control-line reduction scenario into an MXC heat-load estimate.
The estimate combines the RSA-2048 resource estimate of \(381{,}000\) physical qubits from Mundada
et al.~\cite{mundada2026heterogeneous} with the approximate value of three control lines per
qubit inferred from leading experimental systems~\cite{gao2025establishing, he2025experimental,
google2025quantum, sivak2025reinforcement}. The dedicated-line baseline is
\begin{equation}
    N_{\mathrm{ded}}=3\times381{,}000=1{,}143{,}000.
\end{equation}
This is the no-multiplexing reference used below.

For the per-line passive load, we use the publicly documented NbTi version of the Cri/oFlex
flexible stripline as a reference. The product sheet specifies an eight-channel, polyimide-based
stripline with a thickness of approximately \(0.3~\mathrm{mm}\), and reports a low-temperature
passive load of \(350~\mathrm{pW/channel}\) for the NbTi branch in the segment terminating at
the \(10~\mathrm{mK}\) stage~\cite{DelftCircuits2026Brochure}. We therefore use
\begin{equation}
    p_{\mathrm{FPC}}=350~\mathrm{pW/channel}
    =0.350~\mathrm{nW/channel}
\end{equation}
as a manufacturer-reported reference value rather than as a universal measured constant.
The Ag-based value reported in the same product sheet is not used for the present estimate.

Under the idealized frequency assignment shown in Figure~\ref{fig:ideal_frequency_assignment},
the 32-qubit/64-coupler unit gives the multiplexed line count
\begin{equation}
    N_{\mathrm{mux}}=\left\lceil\frac{381{,}000}{32}\right\rceil
    +\left\lceil\frac{2\times381{,}000}{64}\right\rceil
    =23{,}814.
\end{equation}
For comparison, the \(1{:}12\) case contains
\begin{equation}
    N_{1:12}=\frac{1{,}143{,}000}{12}=95{,}250
\end{equation}
shared control paths. The passive MXC load is calculated as
\begin{equation}
    P_{\mathrm{MXC}}=N_{\mathrm{line}}p_{\mathrm{FPC}}.
\end{equation}

For a continuous multiplexing ratio \(R\), the corresponding reference curve is
\begin{equation}
    P_{\mathrm{MXC}}(R)=\frac{N_{\mathrm{ded}}p_{\mathrm{FPC}}}{R}
    =\frac{400.05}{R}~\mu\mathrm{W}.
\end{equation}
The resulting comparison is summarized below.
\begin{center}
    \begingroup
    \scriptsize
    \setlength{\tabcolsep}{2pt}
    \begin{tabular}{@{}lccc@{}}
        \toprule
        Architecture & \(N_{\mathrm{MXC}}\) & \(P_{\mathrm{MXC}}\) & Budget test \\
        \midrule
        Dedicated lines & \(1{,}143{,}000\) & \(400.05~\mu\mathrm{W}\) & \(>30~\mu\mathrm{W}\) \\
        \(1{:}12\) multiplexing & \(95{,}250\) & \(33.34~\mu\mathrm{W}\) & \(>30~\mu\mathrm{W}\) \\
        \(1{:}48\) multiplexing & \(23{,}814\) & \(8.33~\mu\mathrm{W}\) & \(<30~\mu\mathrm{W}\) \\
        \bottomrule
    \end{tabular}
    \endgroup
\end{center}
The continuous curve crosses the representative \(30~\mu\mathrm{W}\) cooling level at
\begin{equation}
    R_{\mathrm{cross}}=\frac{400.05}{30}\approx13.3.
\end{equation}
Thus, under the public NbTi-FPC reference specification, the dedicated-line and \(1{:}12\)
cases exceed the representative MXC cooling level, whereas the idealized \(1{:}48\) case
remains below it~\cite{tian2025high, krinner2019engineering, raicu2025cryogenic}. This comparison
includes only the passive conductive contribution assigned to the control lines; additional loads from filters, switches, splitters, connectors, packaging and other low-temperature components are excluded.

\section{II. Derivation of MP-CZ and the Error Source}
\subsection{\label{appendix_ACCZbase} II.A Basic Derivation of MP-CZ}
We simulate a three-mode qubit--coupler--qubit Hamiltonian with four local levels,
\(\ket{0},\ldots,\ket{3}\), per mode. This defines a formal
\(4^3=64\)-dimensional product basis; for numerical propagation, a reduced
dressed/transition subspace is selected for each waveform and working point, so its
dimension is not fixed a priori. The flux phase \(\phi_i\),
asymmetry \(d_i\), and drive \(u_i(t)\) enter through the effective Josephson energy
\begin{equation}
    \widetilde E_{J,i}(t)=E_{J,ii}^{\mathrm{in}}
    \sqrt{\cos^2(\pi\tilde\phi_i(t))
    +d_i^2\sin^2(\pi\tilde\phi_i(t))} ,
    \label{EJ_flux_map}
\end{equation}
with \(\tilde\phi_i(t)=\phi_i+u_i(t)\).
With \(\lambda_i(t)=[\widetilde E_{J,i}(t)/(8E_{C,ii})]^{1/4}\) and auxiliary
coordinate drives set to zero, the finite-dimensional Hamiltonian is
\begin{equation}
    \begin{aligned}
    H(t)={}&\sum_i\Big\{4E_{C,ii}\left[(\lambda_i(t)\hat p_i)^2-\mathbb I\right]\\
    &\quad-\widetilde E_{J,i}(t)\left[\cos\!\left(\frac{\hat x_i}{\lambda_i(t)}\right)-\mathbb I\right]\Big\}\\
    &+\sum_{i<j}8E_{C,ij}\lambda_i(t)\lambda_j(t)\hat p_i\hat p_j .
    \end{aligned}
    \label{H_numerical_model}
\end{equation}
Here \(\hat x_i=(\hat a_i+\hat a_i^\dagger)/\sqrt{2}\) and
\(\hat p_i=(\hat a_i-\hat a_i^\dagger)/(i\sqrt{2})\). The present diagonal \(E_J\)
input removes an additional Josephson coupling term, while the off-diagonal \(E_C\)
elements provide capacitive coupling. Static spectra follow from diagonalizing
\(H(\phi)\), and gate dynamics from \(i\dot U(t)=H(t)U(t)\).

For the analytical discussion below, the four-level numerical model is projected onto
the lowest three local levels and the explicit coupler-excitation channels are omitted.
This reduced model is used to identify the desired MP-CZ transition and the associated
resonance conditions; it is not the Hamiltonian used for the four-level numerical
propagation:

\begin{equation}
    \begin{split}
    H_{sub}(t)={}&\sum_{kl\in\{10,01,11,20,02\}}\omega_{kl}\ket{k0l}\bra{k0l}\\
    &+ g_{eff}(t)\big(\ket{100}\bra{001}+\sqrt{2}\ket{200}\bra{101}\\
    &+\sqrt{2}\ket{101}\bra{002}+\mathrm{h.c.}\big)
    \end{split}
    \label{H_eff}
\end{equation}

where \(g_{eff}(t)\) is the effective coupling strength of the two qubits, which can be adjusted by the magnetic flux bias of the coupler.
Here \(\omega_{kl}\equiv\omega_{k0l}\) denotes the energy of the three-mode state
\(\ket{k0l}\), with the coupler in \(\ket{0}\).

According to \autoref{H_eff}, if we drive the coupler flux with a microwave tone at \(\omega_d\), \(g_{eff}(t)\) can be modulated and expanded to \(g_{eff}(t)=\sum_{n=-\infty}^{+\infty} {g_{eff}^{(n)} e^{in\omega_dt} }\), where \(g_{eff}^{(n)}\) is determined by the nonlinearity of \(g_{eff}\) shown in \autoref{fig:ACCZ_design}(b). Using the interaction picture, \autoref{H_eff} can be converted to \autoref{HI_eff}.

\begin{equation}
    \begin{split}
    \tilde{H}_{sub,I}(t)= \sum_{n=-\infty}^{+\infty} g_{eff}^{(n)}e^{in\omega_dt}(e^{-i(\omega_{001}-\omega_{100})t}\ket{100}\bra{001}
    \\+e^{-i(\omega_{101}-\omega_{200})t}\sqrt{2}\ket{200}\bra{101}
    \\+e^{-i(\omega_{002}-\omega_{101})t}\sqrt{2}\ket{101}\bra{002} + h.c.)
    \end{split}
    \label{HI_eff}
\end{equation}

For the first-harmonic rotating-wave approximation (RWA), set \(\omega_d = \omega_{101}-\omega_{200} \approx |\eta_1| + \Delta\),
assume a real modulation so that \(g_{eff}^{(-1)}=(g_{eff}^{(1)})^*\), and require
\(|g_{eff}^{(1)}|\ll \bigl|\omega_d-|\omega_{101}-\omega_{002}|\bigr|\) and
\(|g_{eff}^{(1)}|\ll |\omega_d-(\omega_{001}-\omega_{100})|\). Applying the rotating
wave approximation, we obtain:
\begin{equation}
    \begin{split}
        \tilde{H_I} = \sqrt{2}\,g_{eff}^{(1)}\ket{200}\bra{101}
        \\+ \sqrt{2}\,g_{eff}^{(-1)}\ket{101}\bra{200}
    \end{split}
    \label{H_I_11_20}
\end{equation}

\subsection{\label{appendix_errorsour} II.B Error Source of MP-CZ}

Based on the form of the Hamiltonian in the interaction picture given by \autoref{HI_eff}, when we apply the drive pulse, in addition to the desired interaction term in \autoref{H_I_11_20}, there will also be some undesired interaction terms. These undesired interactions can lead to errors. The details are shown in \autoref{H_I}; the analysis is given in \hyperref[g_ic_oscillation]{Coupler Excitation caused by the oscillation of capacitive coupling strength}. 
Potential sources of error include the following.

\begin{enumerate}
    \item \(n\omega_d = \Delta\), causing a \(\ket{001} \leftrightarrow \ket{100}\) swap.
    \item \(n\omega_{d,\mathrm{high}} = \bigl|\Delta - |\eta_2|\bigr|\), causing a \(\ket{101} \leftrightarrow \ket{002}\) swap; \(n\omega_{d,\mathrm{low}} = \Delta + |\eta_1|\), causing a \(\ket{101} \leftrightarrow \ket{200}\) swap.
    \item \(n\omega_d = (\omega_c - \omega_1)\), causing excitation of the coupler.
    \item \(n\omega_d = (\omega_c - \omega_1 - \Delta)\), causing excitation of the coupler.
\end{enumerate}

For high-frequency MP-CZ, \(\omega_{d,\mathrm{high}} = \Delta + |\eta_1|\), the conditions for the first two error sources cannot be satisfied. Therefore, we only need to consider the real excitation leakage to the coupler. For low-frequency MP-CZ, we need to consider all the error sources.

\section{\label{appendix_nonlinear} III. Influence of Driving Amplitude on Coupler Excitation}
\subsection{\label{bessel_analysis} III.A Coupler Excitation caused by high-order Bessel Component }

Regardless of \(g_{ic}\) oscillation, 
we write out the full Hamiltonian for this model:

\begin{equation}
    \begin{split}
    H = \sum_{i=1,2,c}\left(4E_{C,i}n_i^2 - E_{J,i}\cos{\phi_i}\right) 
    \\+ 4e^2\frac{C_{12}}{C_1C_2}n_1n_2 
    \\+ \sum_{i=1,2}4e^2\frac{C_{ic}}{C_i C_c}n_i n_c 
    \end{split}
\end{equation}

For the analytical error analysis, we project the four-level numerical model onto the
lowest three local levels and obtain the following reduced Hamiltonian:

\begin{equation}
    \begin{split}
    H = \omega_1\ket{1}_1\bra{1} + (2\omega_1 - |\eta_1|)\ket{2}_1\bra{2} 
    \\+ \omega_c\ket{1}_c\bra{1} + (2\omega_c - |\eta_c|)\ket{2}_c\bra{2} \\
    \quad + \omega_2\ket{1}_2\bra{1} + (2\omega_2 - |\eta_2|)\ket{2}_2\bra{2} \\
    \quad + g_{12}(\ket{10}\bra{01} + \sqrt{2}\ket{20}\bra{11} \\+ \sqrt{2}\ket{11}\bra{02} + \text{h.c.})_{12} \otimes I_c \\
    \quad + g_{1c}(\ket{10}\bra{01} + \sqrt{2}\ket{20}\bra{11} \\+ \sqrt{2}\ket{11}\bra{02} + \text{h.c.})_{1c} \otimes I_2  \\
    \quad + g_{c2}(\ket{10}\bra{01} + \sqrt{2}\ket{20}\bra{11} \\+ \sqrt{2}\ket{11}\bra{02} + \text{h.c.})_{c2} \otimes I_1 
    \end{split}
\end{equation}

Here, \(\omega_i\) is the energy difference between \(\ket{0}\) and \(\ket{1}\) for the \(i\)-th qubit, and \(|\eta_i|\) represents the anharmonicity of the qubit. The coupling strength between qubits \(i\) and \(j\) is denoted by \(g_{ij}\), where \(i, j=1,2,c\) includes the coupler.

Injecting a flux pulse into the coupler modulates the effective coupling strength.

\begin{equation}
    \begin{split}
    \omega_c=\omega_{c,\max}\sqrt{|\cos{\frac{\pi\Phi_e}{\Phi_0}}|}
    \\=\omega_{c,\max}\sqrt{|\cos{\phi(t)}|}
    \\=\omega_{c,\max}\sqrt{|\cos{(\phi_0+\epsilon_\phi \cos{(\omega_d t+\theta_d)})}|}
    \end{split}
\end{equation}

Retaining the DC and fundamental Fourier components, we approximate:
\(\omega_c(t)\simeq\overline{\omega_c}+\epsilon*\cos{(\omega_d t+\theta_d)}\),
where \(\overline{\omega_c}\) is the cycle-averaged coupler frequency at the given modulation amplitude.

By converting to the interaction picture and using the methods and conclusions from \cite{didier2018analytical, reagor2018demonstration}, we can obtain:
\begin{equation}
    \begin{aligned}
    \tilde{H}_I(t)={}&g_{12}\Big(e^{-i(\omega_2-\omega_1)t}\ket{10}\bra{01}\\
    &\quad+\sqrt{2}e^{-i(\omega_2-\omega_1+|\eta_1|)t}\ket{20}\bra{11}\\
    &\quad+\sqrt{2}e^{-i(\omega_2-\omega_1-|\eta_2|)t}\ket{11}\bra{02}\Big)_{12}\otimes I_c+\mathrm{h.c.}\\
    &+\sum_{n=-\infty}^{+\infty}g_{1c}^{(n)}e^{in\omega_dt}
    \Big(e^{-i(\omega_c-\omega_1)t}\ket{10}\bra{01}\\
    &\quad+\sqrt{2}e^{-i(\omega_c-\omega_1+|\eta_1|)t}\ket{20}\bra{11}\\
    &\quad+\sqrt{2}e^{-i(\omega_c-\omega_1-|\eta_c|)t}\ket{11}\bra{02}\Big)_{1c}\otimes I_2+\mathrm{h.c.}\\
    &+\sum_{n=-\infty}^{+\infty}g_{c2}^{(n)}e^{in\omega_dt}
    \Big(e^{-i(\omega_2-\omega_c)t}\ket{10}\bra{01}\\
    &\quad+\sqrt{2}e^{-i(\omega_2-\omega_c+|\eta_c|)t}\ket{20}\bra{11}\\
    &\quad+\sqrt{2}e^{-i(\omega_2-\omega_c-|\eta_2|)t}\ket{11}\bra{02}\Big)_{c2}\otimes I_1+\mathrm{h.c.}
    \end{aligned}
    \label{H_I}
\end{equation}

For the modulated couplings \(ij\in\{1c,c2\}\),
\(g_{ij}^{(n)}=g_{ij}J_n(\epsilon/\omega_d)e^{i\beta_{ij}^{(n)}}\), where \(J_n\) is the
Bessel function of the first kind of order \(n\) and the interaction phases of the two
channels are
\[
\beta_{1c}^{(n)}=n(\theta_d+\pi)+\frac{\epsilon}{\omega_d}\sin\theta_d,
\qquad
\beta_{c2}^{(n)}=n\theta_d-\frac{\epsilon}{\omega_d}\sin\theta_d.
\]

According to \autoref{H_I}, the corresponding coupling or excitation amplitude is
\(|g_{ij}^{(n)}|=|g_{ij}||J_n(\epsilon/\omega_d)|\).
Since \(g_{ij}\) is fixed, the relevant quantity is \(|J_n(\epsilon/\omega_d)|\).

\begin{figure}
    \centering 
    \includegraphics[width=0.7\linewidth]{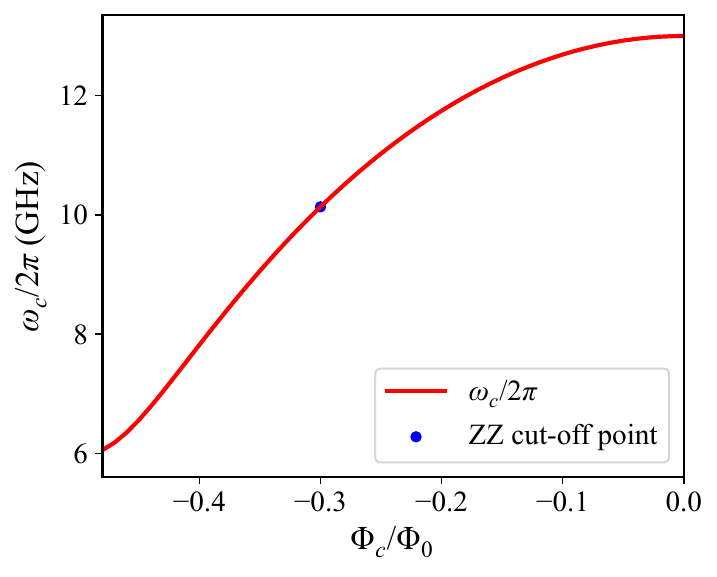}
    \caption{The relationship between the \(\omega_c\) of the coupler and the magnetic flux bias of the coupler, with blue circles indicating the coupling zero point}
    \label{f01coupler_Phi}
\end{figure}

\begin{figure}
    \centering
    \includegraphics[width=0.7\linewidth]{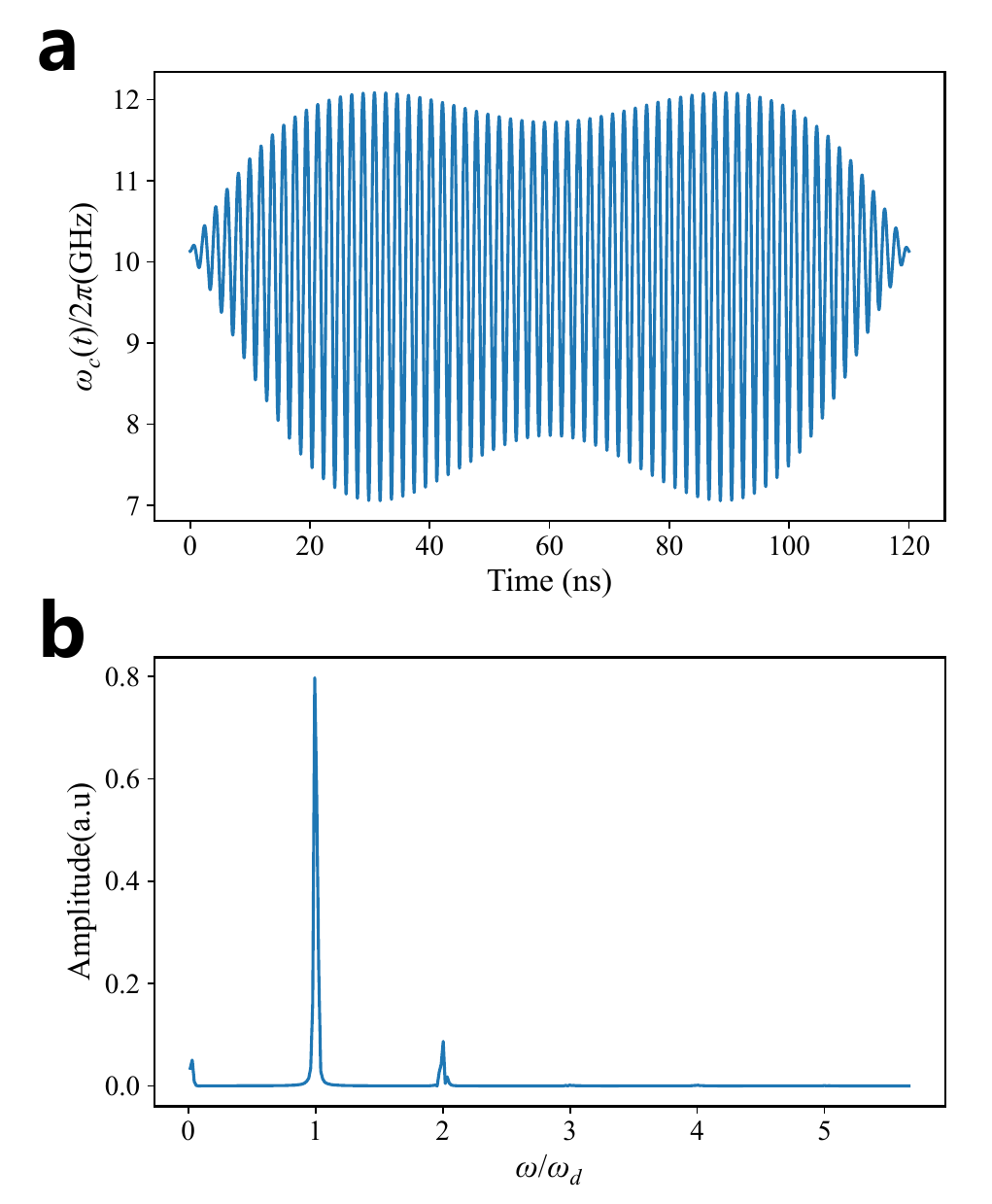}
    \caption{(a) Evolution of \(\omega_c(t)/2\pi\). (b) Its spectrum.}
    \label{f01Coupler_evolution}
\end{figure}

The coupler frequency depends strongly and nonlinearly on the coupler flux bias, as shown in \autoref{f01coupler_Phi}. 
When the coupler flux bias is driven by a microwave tone, the evolution of \(\omega_c/2\pi\) is shown in \autoref{f01Coupler_evolution}, in which \(\omega_d/(2\pi)=307.15~\mathrm{MHz}\) and the modulation amplitude is 0.119\(\Phi_0\). Because \(\omega_c/2\pi\) depends nonlinearly on the coupler flux bias, \autoref{f01Coupler_evolution}(b) shows multiple frequencies. 

In \autoref{H_I}, we considered only the first order of \(\omega_c\), since \(\omega_c(t)=\overline{\omega_c}+\epsilon*\cos{(\omega_d t+\theta_d)}\).
Now we consider \(\omega_c(t)=\overline{\omega_c}+\epsilon_1*\cos{(\omega_d t+\theta_{d,1})} + \epsilon_2*\cos{(2\omega_d t+\theta_{d,2})}\)
Two factors must therefore be taken into account: \(|J_n(\epsilon_1/\omega_d)|\) and \(|J_m(\epsilon_2/2\omega_d)|\), where \( m \sim n/2\). The choice \(m\sim n/2\) is used for an order-of-magnitude comparison at comparable sideband frequencies. The \(\epsilon_1\) and \(\epsilon_2\) can be calculated using \(\omega_c(t)\).

\begin{figure}
    \centering
    \includegraphics[width=0.7\linewidth]{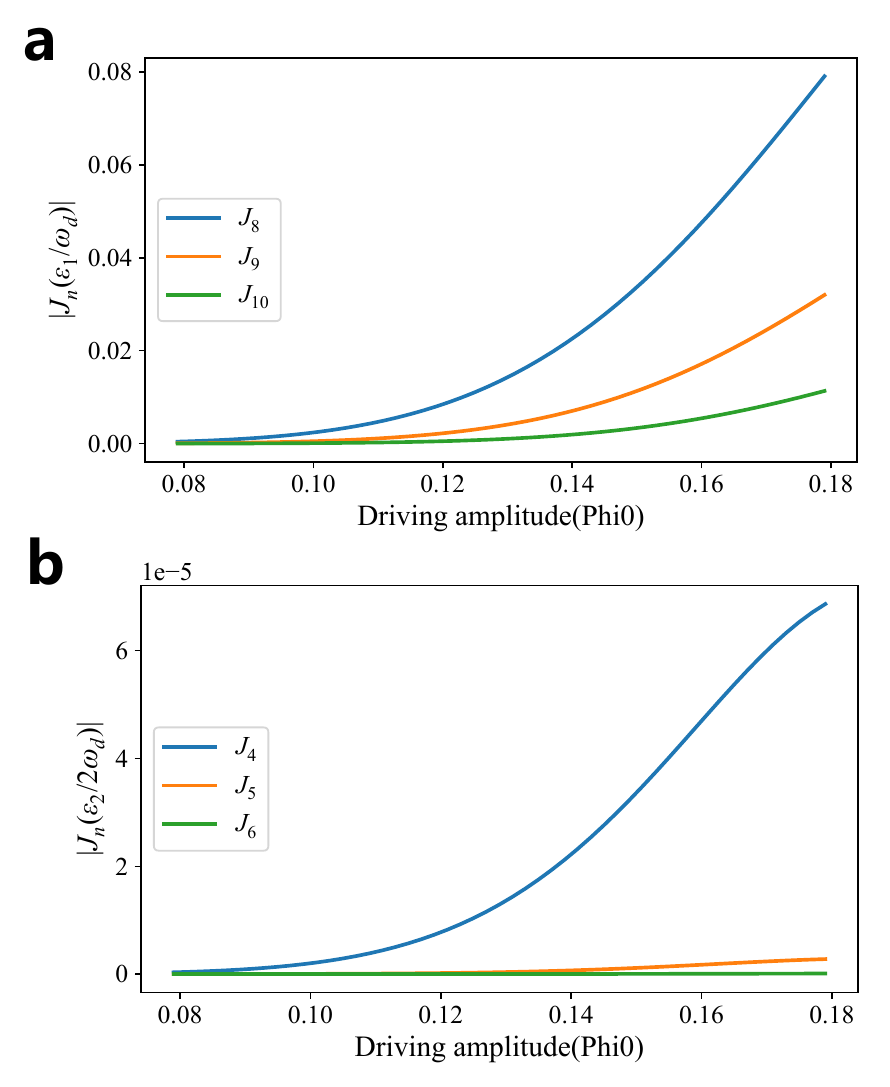}
    \caption{Bessel function values at several orders versus driving amplitude. (a) First-order effect. (b) Second-order effect.}
    \label{BesselJ_amps}
\end{figure}

Holding the driving frequency fixed and sweeping the driving amplitude, we evaluate both factors and obtain \autoref{BesselJ_amps}. 
The Bessel function values increase with driving amplitude, and coupler excitation becomes correspondingly more severe. 
In addition, the evaluated Bessel factors associated with the second-harmonic modulation are much smaller than those associated with the fundamental modulation over the considered amplitude range, suggesting that the single-harmonic approximation used in \autoref{H_I} is reasonable. 

\subsection{\label{g_ic_oscillation} III.B Coupler Excitation caused by the oscillation of capacitive coupling strength }
\begin{figure}
    \centering
    \includegraphics[scale=0.3]{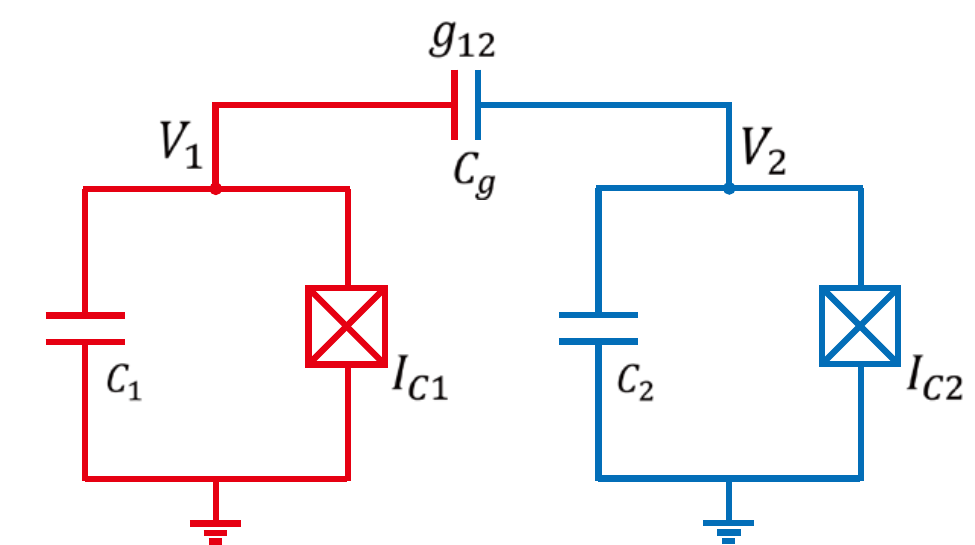}
    \caption{Capacitive coupling of two transmon qubits}
    \label{fig:capacitive_coupling}
\end{figure}

First, we consider the capacitive coupling of two transmon qubits shown in \autoref{fig:capacitive_coupling}. From the basic Lagrangian, we can derive the Hamiltonian of the system and obtain the coupling strength between the two transmons. The result is shown in \autoref{capacitive_coupling}.
The coupling strength is then found to be proportional to \(\sqrt{\omega_1 \omega_2}\). 

In the QCQ model in \autoref{fig:ACCZ_design}(a), we need to modulate \(\omega_c\) to realize the CZ gate, which may cause oscillations of \(g_{1c}\) and \(g_{c2}\). According to \autoref{capacitive_coupling}, we can approximately get \(g_{1c}(t)=g_{1c,idle}\sqrt{\frac{\omega_c(t)}{\omega_{c,idle}}}\), \(g_{c2}(t)=g_{c2,idle}\sqrt{\frac{\omega_c(t)}{\omega_{c,idle}}}\). In other words, \(g_{ic}=g_{ic,idle}*\sqrt{\frac{1}{\omega_{c,idle}}(\overline{\omega_c}+\sum_{n=1}^{+\infty}\epsilon_n \cos{(n\omega_dt +\theta_n)})} \). Due to the nonlinearity of the square root, as well as the nonlinearity of \(\omega_c(\Phi_c)\), there will be multiple-frequency components in the evolution of \(g_{ic}\). When \(n\omega_d\approx \overline{\omega_c}-\overline{\omega_{1}}\) or \(n\omega_d\approx \overline{\omega_c}-\overline{\omega_{2}}\), excitation of the coupler may occur. 

\begin{figure}
    \centering
    \includegraphics[width=0.7\linewidth]{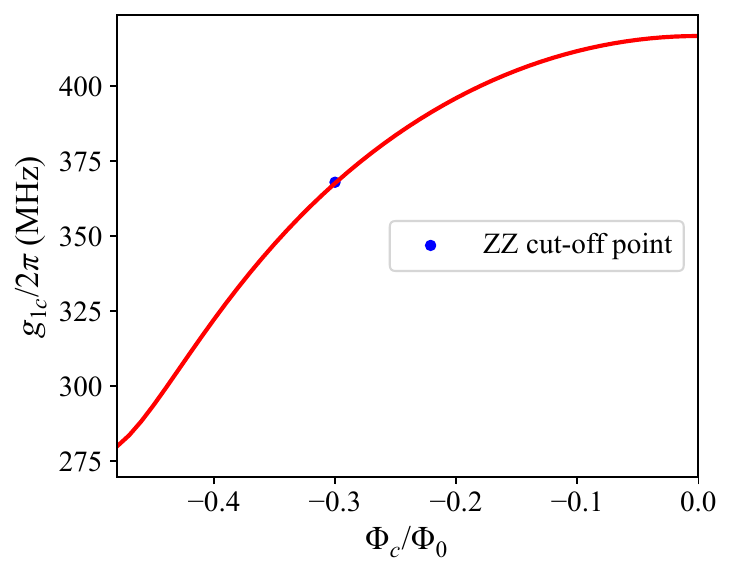}
    \caption{Capacitive coupling strength of qubit1 and the coupler's dependence on the flux bias on the coupler in QCQ system, with blue circles indicating the coupling zero point. }
    \label{fig:g1c}
\end{figure}

\begin{figure*}
    \centering
    \includegraphics[width=0.75\linewidth]{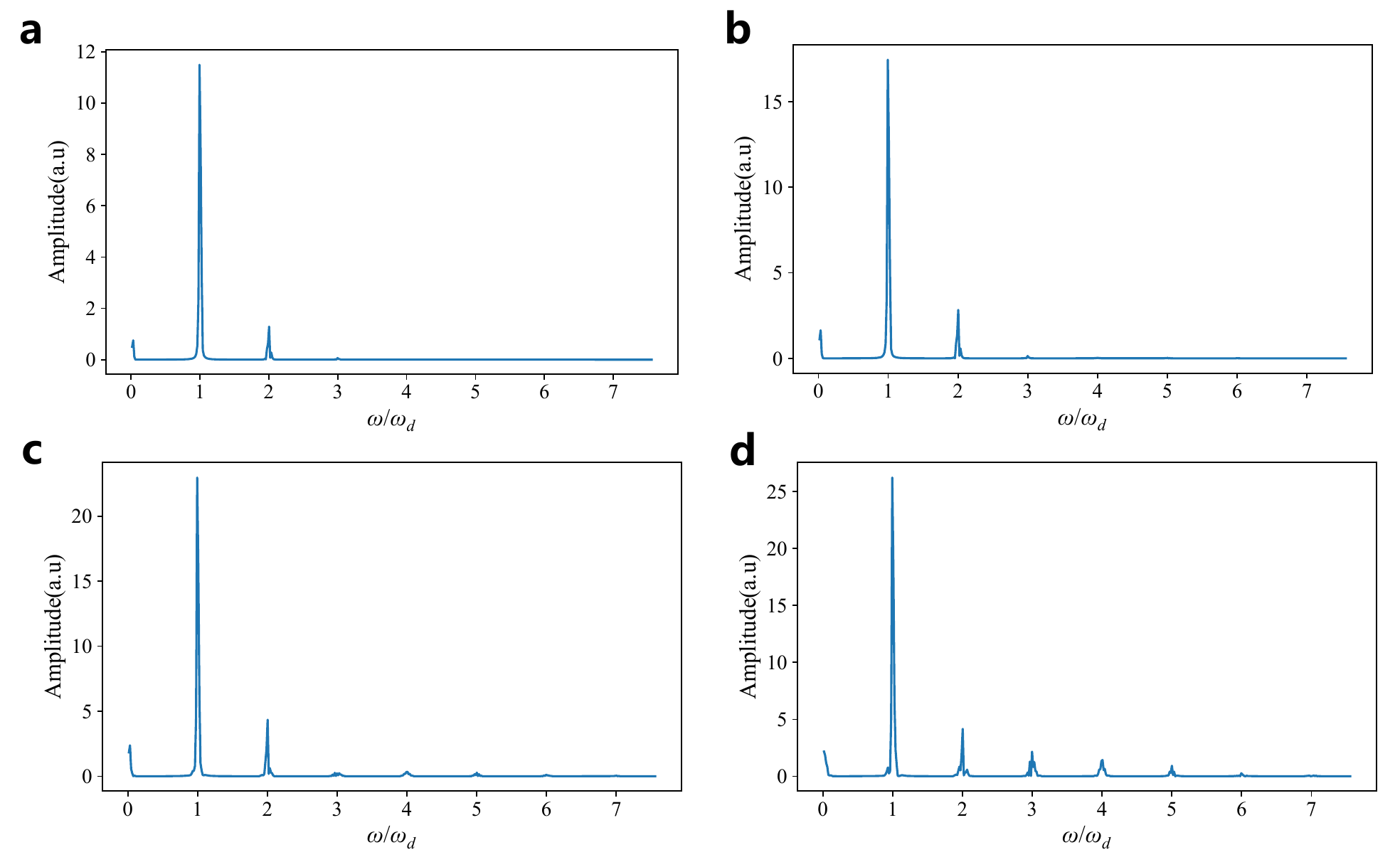}
    \caption{Spectrum of \(g_{1c}(t)\) at different driving amplitudes. (a) amplitude=0.1\(\Phi_0\); (b) amplitude=0.15\(\Phi_0\); (c) amplitude=0.2\(\Phi_0\); (d) amplitude=0.25\(\Phi_0\)}
    \label{fig:g1c_evo}
\end{figure*}

Because an analytical result is difficult to derive, we perform a numerical simulation. Taking \(g_{1c}\) as an example, we calculate the spectrum of \(g_{1c}(t)\) at different driving amplitudes. The result is shown in \autoref{fig:g1c_evo}. The simulation uses the parameters listed in \autoref{simu_paras}. 
The simulations show that the high-order frequency components grow rapidly with driving amplitude, which leads to coupler excitation. This sets a limitation on the MP-CZ gate length. 

\begin{figure*}
\begin{equation}
\begin{split}
L&=\frac{1}{2}C_1\dot{\Phi}_1^2+\frac{1}{2}C_2\dot{\Phi}_2^2+\frac{1}{2}C_g(\dot{\Phi}_1-\dot{\Phi}_2)^2
-E_{J1}(1-\cos{\phi_1})-E_{J2}(1-\cos{\phi_2})\\
Q_1&=\frac{\partial L}{\partial\dot\Phi_1}=(C_1+C_g)\dot{\Phi}_1-C_g\dot{\Phi}_2, 
Q_2=\frac{\partial L}{\partial\dot\Phi_2}=-C_g\dot\Phi_1+(C_2+C_g)\dot\Phi_2\\
H&=\frac{(C_2+C_g)Q_1^2+(C_1+C_g)Q_2^2+2C_gQ_1Q_2}{2(C_1C_2+C_1C_g+C_2C_g)}
+E_{J1}(\frac{1}{2}\phi_1^2-\frac{1}{24}\phi_1^4)+E_{J2}(\frac{1}{2}\phi_2^2-\frac{1}{24}\phi_2^4)\\
&=\frac{(C_2+C_g)Q_1^2+(C_1+C_g)Q_2^2+2C_gQ_1Q_2}{2(C_1C_2+C_1C_g+C_2C_g)}
+2I_{C1}\frac{\Phi_0}{2\pi}\cos{\frac{\pi\Phi_{C1}}{\Phi_0}}(\frac{1}{2}\phi_1^2-\frac{1}{24}\phi_1^4)
+2I_{C2}\frac{\Phi_0}{2\pi}\cos{\frac{\pi\Phi_{C2}}{\Phi_0}}(\frac{1}{2}\phi_2^2-\frac{1}{24}\phi_2^4)\\
&=\hbar\omega_1a_1^{\dagger}a_1+\frac{1}{2}\hbar\alpha_1 a_1^{\dagger}a_1^{\dagger}a_1a_1
+\hbar\omega_2a^{\dagger}_2a_2+\frac{1}{2}\hbar\alpha_2 a_2^{\dagger}a_2^{\dagger}a_2a_2
-\hbar g(a_1^{\dagger}-a_1)(a_2^{\dagger}-a_2)\\
\omega_1&=\sqrt{\frac{E_{J1}}{(\Phi_0/2\pi)^2C_1}}, 
\omega_2=\sqrt{\frac{E_{J2}}{(\Phi_0/2\pi)^2C_2}}, 
\omega_1=\sqrt{\frac{4\pi I_{C1} \cos{\frac{\pi\Phi_{C1}}{\Phi_0}}}{\Phi_0C_1}}, 
\omega_2=\sqrt{\frac{4\pi I_{C2} \cos{\frac{\pi\Phi_{C2}}{\Phi_0}}}{\Phi_0C_2}},
g=\frac{C_g\sqrt{\omega_1\omega_2}}{2\sqrt{(C_1+C_g)(C_2+C_g)}}
\end{split}
\label{capacitive_coupling}
\end{equation}
\end{figure*}

\section{\label{appendix_coupler_exci} IV. Influence of Driving Frequency on Coupler Excitation}
According to the analysis in \hyperref[appendix_nonlinear]{Supplementary Section~III}, we know that the coupler will be excited when multiple-frequency components of the driving pulse satisfy the following conditions:
\begin{equation}
    \begin{split}
        \\ n\omega_d=\omega_{lc1}=\omega_c-\omega_1
        \\ n\omega_d=\omega_{lc2}=\omega_c-\omega_1-\Delta
    \end{split}
    \label{coupler_excitation_analysis0}
\end{equation}
Since the driving frequency is equal to \(\Delta+|\eta_1|\), we obtain:
\begin{equation}
    \begin{split}
        \\ \text{let } n\omega_d=\omega_{lc1} 
        \\\Rightarrow n(\Delta+|\eta_1|)=\omega_c-\omega_1 
        \\\Rightarrow n\Delta=\omega_c-\omega_1-n|\eta_1| 
        \\\Rightarrow \Delta_{lc1}=\frac{\omega_c-\omega_1}{n}-|\eta_1|
        \\
        \\ \text{let } n\omega_d=\omega_{lc2} 
        \\\Rightarrow n(\Delta+|\eta_1|)=\omega_c-\omega_1-\Delta 
        \\\Rightarrow n\Delta=\omega_c-\omega_1-n|\eta_1|-\Delta 
        \\\Rightarrow \Delta_{lc2}=\frac{\omega_c-\omega_1}{n+1}-\frac{n}{n+1}|\eta_1|
    \end{split}
    \label{coupler_excitation_analysis}
\end{equation}
Therefore, the coupler will be excited when the frequency difference between the qubits is \(\Delta=\frac{\omega_c-\omega_1}{n}-|\eta_1|\) or \(\Delta=\frac{\omega_c-\omega_1}{n+1}-\frac{n}{n+1}|\eta_1|\).

\section{\label{appendix_fdm} V. Error from MP-CZ FDM}

FDM introduces two main types of infidelity: crosstalk and waveform distortion. 

Crosstalk-induced infidelity is shown in \autoref{fig:Harmonic_and_FDM}(b,c), which is simulated using parameters in \autoref{simu_paras}. 
\autoref{fig:Harmonic_and_FDM}(c) is simulated using Qubit 1 from \autoref{simu_paras}. 
For MP-CZ, if the frequency spacing of the FDM is 100~MHz, an isolation of 50~dB is required. For single-qubit \(XY(\pi/2)\) gates, if the frequency spacing is 100~MHz, an isolation of 10~dB is required. 
In our experiment, these conditions are easily met. 

Pulse-distortion-induced infidelity arises from the variation in group delay within the pass-band of filters. For MP-CZ, the distortion does not directly degrade fidelity. Instead, it degrades the subsequent single-qubit gates in XEB or QEC circuits. 
We can address the distortion by offsetting the variation in group delay at the software level. The target is to minimize the residual waveform on the Z-channel of couplers within the duration of the single-qubit gate. 
The distortion and precorrected waveforms used in our experiment are shown in \autoref{fig:Harmonic_and_FDM}(d-k); the correction uses an FIR pre-distortion method based on the S21 response measured by room-temperature vector network analysis. 

\section{\label{pre_cali} VI. Calibration of single-qubit gates}

\subsection{\label{gd_bpf} VI.A Group delay in the band-pass filter}

Due to the group delay characteristics of the band-pass filter (BPF), microwave waveforms with envelopes experience both temporal delay and signal trailing after passing through the filter.

\begin{figure}[htbp]
    \centering
    \includegraphics[width=\linewidth]{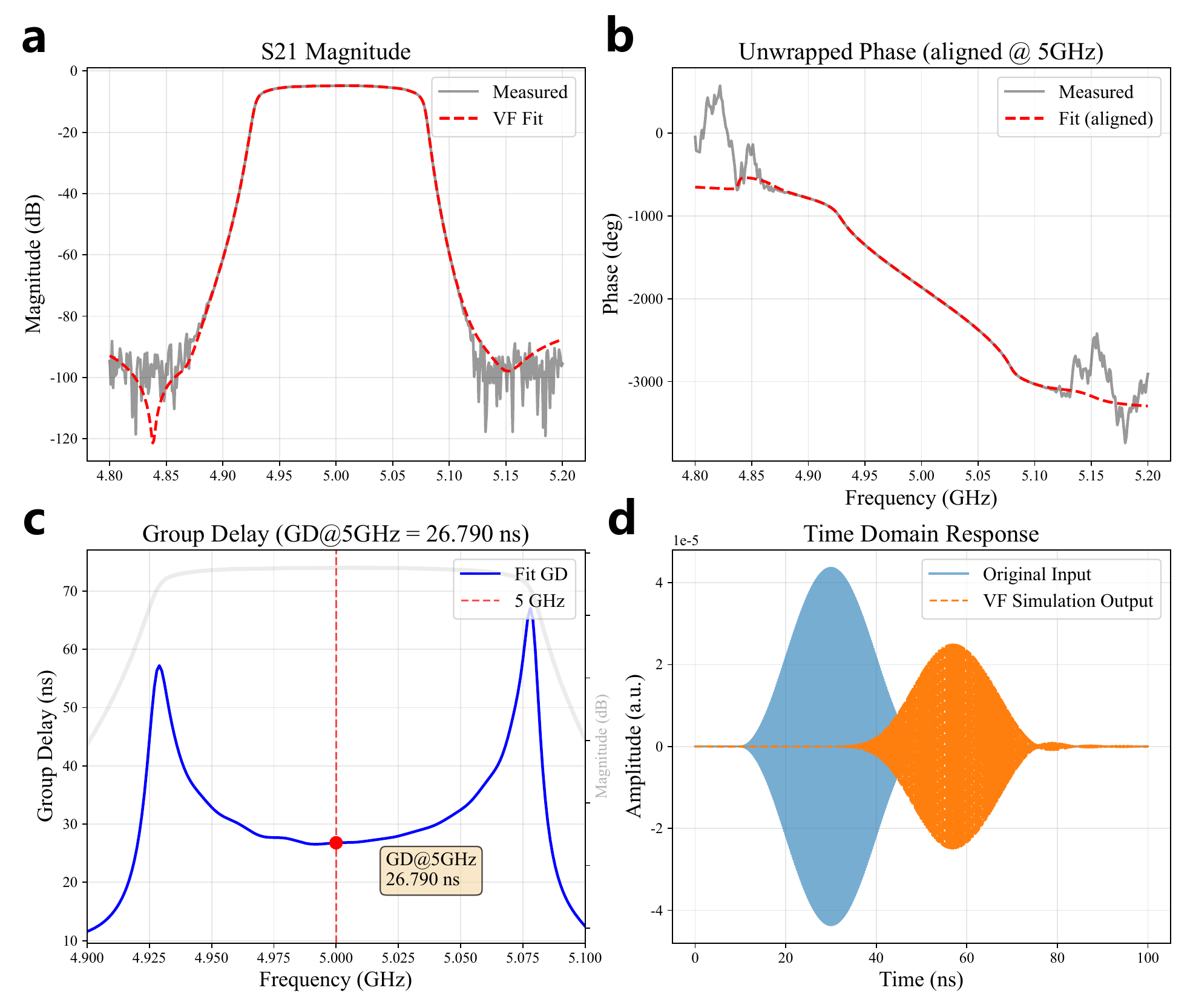}
    \caption{
    Vector fitting results of the band-pass filter used in the experiment. (a) Measured and fitted $S_{21}$ magnitude. (b) Unwrapped phase response. (c) Group delay. (d) Original and filtered time-domain waveforms. 
    }
    \label{fig:vf_result}
\end{figure}

In \autoref{fig:vf_result}, we characterize the transmission properties of the room-temperature BPF and extract the group delay using vector fitting. Based on this analysis, we simulate the temporal profile of a carrier signal with a Hanning envelope after propagating through the BPF. The simulation predicts a group delay of approximately 26.8 ns, along with characteristic pulse broadening and trailing effects induced by the filter's frequency-dependent phase response.

\begin{figure}[htbp]
    \centering
    \includegraphics[width=0.7\linewidth]{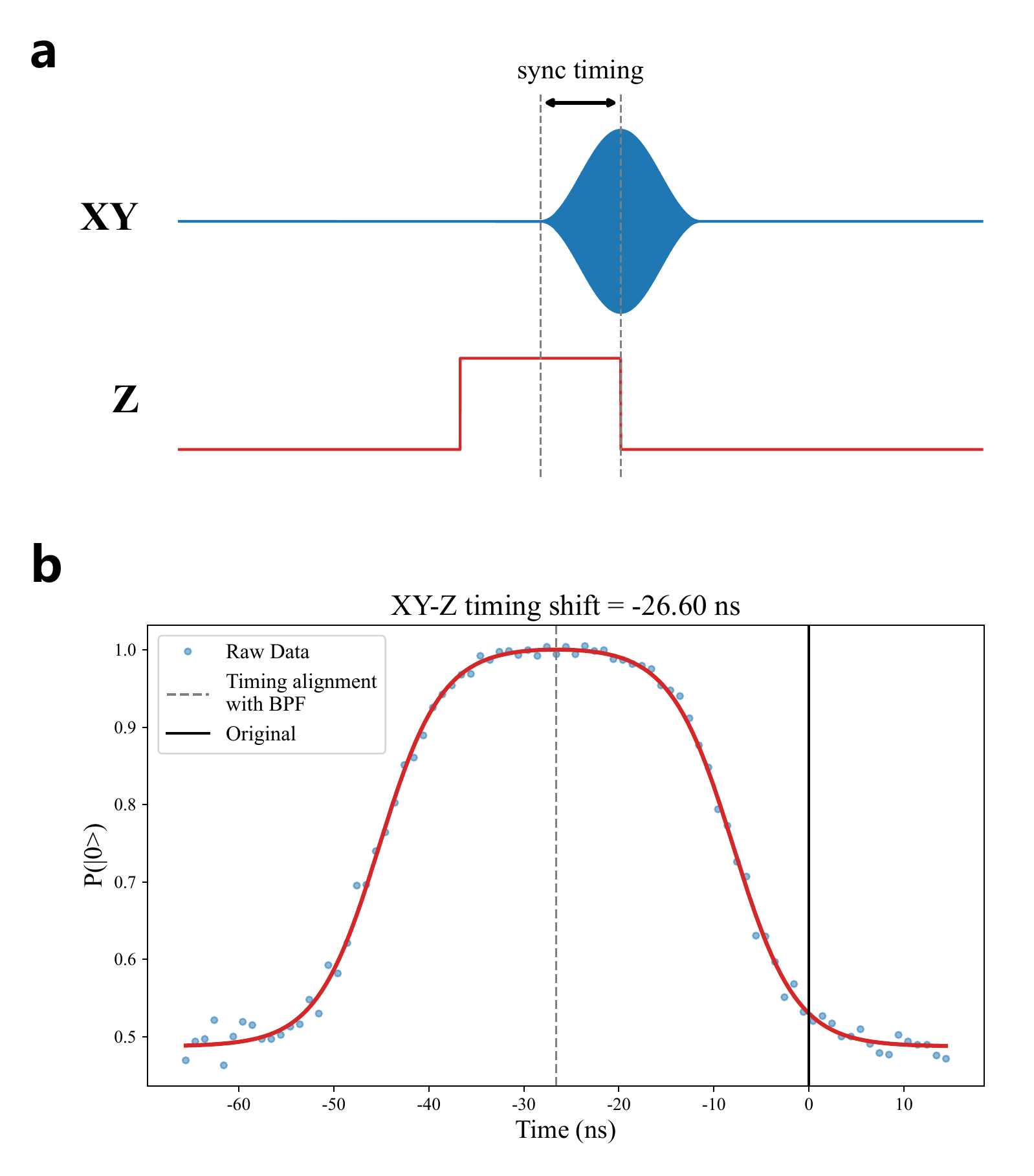}
    \caption{(a) Circuit scheme for XY-Z timing calibration. The $X_{\pi/2}$ gate is calibrated at the idle point. By scanning the synchronization timing, the $X_{\pi/2}$ gate fails when the square pulse overlaps with it, resulting in $P_{\ket{0}}=1$ when the pulse overlaps and $P_{\ket{0}}=0.5$ otherwise, which identifies the alignment point. (b) Experimental XY-Z channel timing difference. For clarity, we show only the data from Qubit 3, whose configuration matches the 5 GHz BPF simulated in \autoref{fig:vf_result}. }
    \label{fig:xyz_timing}
\end{figure}

To experimentally verify this prediction, we measure the timing shift between the XY and Z control channels upon inserting the BPF into the signal path, using circuits in \autoref{fig:xyz_timing}~(a). As shown in \autoref{fig:xyz_timing}~(b), the XY channel exhibits a timing delay of approximately 26.60 ns relative to the Z channel, which is in excellent agreement with the group delay calculated from the transmission measurement in \autoref{fig:vf_result}. This consistency between simulation and experiment confirms that the observed timing shift originates from the BPF's group delay characteristics.

\subsection{\label{fir_predistortion} VI.B FIR pre-distortion for single-qubit FDM}

To compensate for the distortion introduced by the band-pass filter (BPF) on the in-band $X_{\pi/2}$ gate waveform, we implement a pre-distortion scheme on the control waveform to suppress signal trailing and reduce group delay. Specifically, we first characterize the transmission properties of the BPF at room temperature and extract its S21 scattering parameters. The kernel of a finite-impulse-response (FIR) filter is then derived by inverting the frequency response function obtained from S21. The pre-distorted waveform is generated by convolving the original input waveform with this FIR kernel. 

\begin{figure*}[!htbp]
    \centering 
    \includegraphics[width=0.85\linewidth]{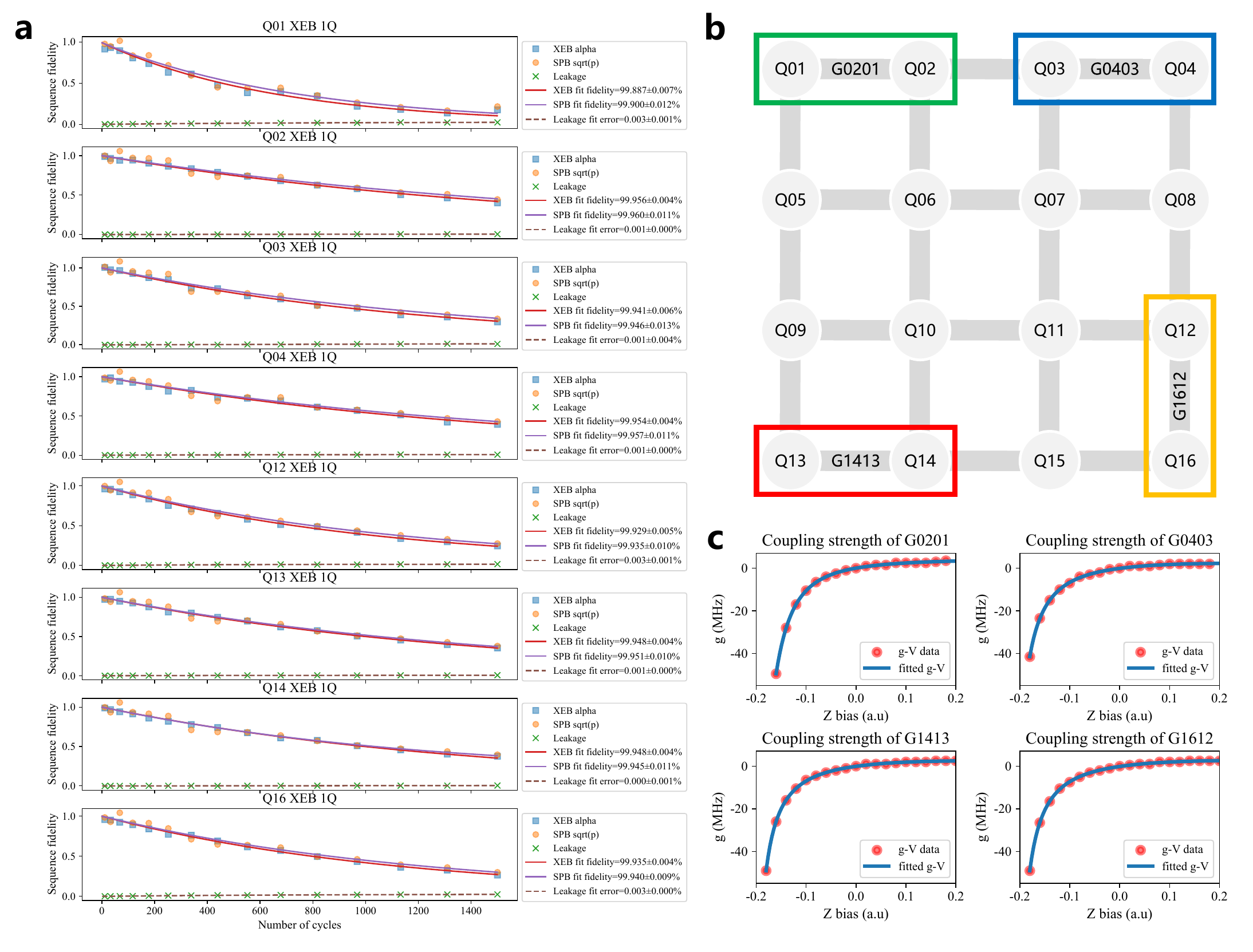}
    \caption{
    (a) XEB results for the \(\pi/2\) single-qubit Pauli gate, collected from the eight qubits. 
    (b) Topology of the qubits and couplers. The four couplers used here are marked with boxes. 
    (c) Coupling strength versus coupler Z bias. 
    }
    \label{Sfig:exp_and_S21Z}
\end{figure*}

\begin{table*}[htbp]
    \centering
    \caption{Parameters of the qubits used in the MP-CZ experiment}
    \begin{tabular}{ccccccccc}
        \toprule
        Parameter & Q01 & Q02 & Q03 & Q04 & Q12 & Q13 & Q14 & Q16 \\
        \midrule
        \(f_{01,\max}(\text{GHz})\) & 5.568 & 4.869 & 4.942 & 4.941 & 4.919 & 4.82 & 5.257 & 5.426 \\
        \(f_{01}(\text{GHz})\) & 5.344 & 4.853 & 4.616 & 4.857 & 4.803 & 4.793 & 5.145 & 5.124 \\
        \(\eta(\text{MHz})\) & -183 & -189 & -188 & -190 & -189 & -187 & -185 & -186 \\
        \(X_{\pi/2}\ error\) & \(1.1\times10^{-3}\) & \(4.5\times10^{-4}\) & \(6.2\times10^{-4}\) & \(4.6\times10^{-4}\) & \(7.1\times10^{-4}\) & \(5.2\times10^{-4}\) & \(5.2\times10^{-4}\) & \(6.5\times10^{-4}\)\\
        \(T_1\ \mathrm{idle}\ (\mu\mathrm{s})\) & 50 & 63 & 73 & 59 & 25 & 65 & 65 & 50 \\
        \bottomrule
    \end{tabular}
    \label{exp_parameters}
\end{table*}

\section{\label{appendix_exp_parameters} VII. Experimental Conditions}

\subsection{\label{appendix_exp_parameters_2q} VII.A Experimental Conditions for MP-CZ Gates}
The topology of the qubits and the couplers is shown in 
\autoref{Sfig:exp_and_S21Z}(b).
In our system, Q05 cannot be driven due to a line fault.
Q07's \(f01_{\mathrm{max}}\) is 4.38~GHz, which is too low to perform MP-CZ. 
Q08 and Q11 are not stable in $T_{\phi}$.
Accounting for these factors, we select G0201, G0403, G1413 and G1612, together with their qubits, for the MP-CZ FDM experiments. 
G0403, G1612, G0201 and G1413 are referred to as Coupler 1, Coupler 2, Coupler 3 and Coupler 4, respectively. 
The parameters of the qubits are shown in 
\autoref{exp_parameters}.
The relation between coupling strength and coupler Z bias is shown in \autoref{Sfig:exp_and_S21Z}(c).

Two observations regarding the experimental conditions are worth noting. 
In Cond.~1, a flat-top envelope performs better than the envelope shown in \autoref{fig:Harmonic_and_FDM}(a) for Couplers 2–4, so the flat-top envelope is used for these couplers in this condition. 
In Cond.~2–4, where the signal has a narrower spectral width, the envelope shown in \autoref{fig:Harmonic_and_FDM}(a) performs better, so this envelope is used in these conditions.

\subsection{\label{appendix_exp_parameters_1q} VII.B Experimental Conditions for Single-qubit Gates}
In this experiment, we selected four qubits from those used in the MP-CZ experiment. Between the two experiments, the dilution refrigerator underwent one thermal cycle, resulting in slight variations in qubit performance.

Q01, Q02, Q03 and Q04 are referred to as Qubit 1, Qubit 2, Qubit 3 and Qubit 4, respectively. 
The parameters of the qubits during the experiment are shown in 
\autoref{exp_parameters_1q}.

\begin{table}[htbp]
    \centering
    \caption{Parameters of the qubits used in the single-qubit gate experiment}
    \begin{tabular}{ccccccccc}
        \toprule
        Parameter & Q01 & Q02 & Q03 & Q04\\
        \midrule
        \(f_{01}(GHz)\) & 5.415 & 4.752 & 4.846 & 4.957  \\
        \(T_1\ \mathrm{idle}\ (\mu\mathrm{s})\) & 37 & 63 & 61 & 62 \\
        \bottomrule
    \end{tabular}
    \label{exp_parameters_1q}
\end{table}

\section{\label{appendix_simulation_parameters} VIII. Numerical Simulation Model and Parameters}
For reproducibility, the runtime parameter set for the numerical MP-CZ simulations is
given below. The three-mode QCQ model retains four local levels,
\(\ket{0}\) through \(\ket{3}\); the actual propagation uses a reduced
dressed/transition subspace selected for each waveform and working point. The
effective capacitance-energy matrix before the runtime \(2\pi\) scaling is
\begin{equation}
E_C=\begin{pmatrix}
0.18490 & 0.01600 & 0.002295\\
0.01600 & 0.14190 & 0.014908\\
0.002295 & 0.014908 & 0.18569
\end{pmatrix},
\end{equation}
where the off-diagonal entries are kept as supplied; the coupling term uses the
upper-triangular \(E_{C,ij}\) entries for \(i<j\), without additional symmetrization.
The runtime input is \(2\pi E_C\). The Josephson-energy input is
\begin{equation}
E_J=\operatorname{diag}(137.132969,\,951.864137,\,120.758883),
\end{equation}
where the displayed values are rounded for readability and already include \(2\pi\);
no additional \(2\pi\) factor is applied to this matrix.
The matrices follow the internal frequency-energy convention of the numerical model; the
remaining dimensionless settings are \(d=[0,0.2,0]\) and \(\phi=[0,-0.293,0]\).
The coupler anharmonicity in Table~I is extracted from the diagonalized model, while
\(g_{12}\), \(g_{1c}\), and \(g_{c2}\) are extracted from its dressed spectrum.

\end{document}